\documentclass[prl,twocolumn,preprintnumbers,longbibliography,amsmath,amssymb,superscriptaddress,showkeys]{revtex4-1}
\usepackage{graphicx,subfigure,amsmath,bbm}
\usepackage{float}
\RequirePackage{import}
\usepackage{graphicx}
\usepackage{dcolumn}
\usepackage{bm}
\usepackage{float}
\usepackage{hyphenat}
\usepackage{natbib}
\usepackage{hyperref}
\usepackage[dvipsnames]{xcolor}
\usepackage{physics}
\usepackage{amsmath} 

\newcommand{\heriotwatt}{Institute of Photonics and Quantum Sciences, SUPA, Heriot-Watt University, Edinburgh EH14 4AS, UK}
\newcommand{\TUM}{Walter Schottky Institute and Physics Department, Technical University of Munich, Am Coulombwall 4a, 85748 Garching, Germany}
\newcommand{\Munster}{{Institute of Physics, University of Münster, Wilhelm-Klemm-Str. 10, 48149 Münster, Germany}}

\begin{document}

\title{Interlayer and moir{\'e} excitons in atomically thin double layers: from individual quantum emitters to degenerate ensembles}

\author{Mauro Brotons-Gisbert}
\email{m.brotons\_i\_gisbert@hw.ac.uk}
\affiliation{\heriotwatt} 
\author{Brian D. Gerardot}
\email{B.D.Gerardot@hw.ac.uk}
\affiliation{\heriotwatt}
\author{Alexander W. Holleitner}
\email{holleitner@wsi.tum.de}
\affiliation{\TUM} 
\author{Ursula Wurstbauer}
\email{wurstbauer@uni-muenster.de}
\affiliation{\Munster} 
    
\date{\today}

\begin{abstract}
Interlayer excitons (IXs), composed of electron and hole states localized in different layers, excel in bilayers composed of atomically thin van der Waals materials such as semiconducting transition metal dichalcogenides (TMDs) due to drastically enlarged exciton binding energies, exciting spin-valley properties, elongated lifetimes, and large permanent dipoles. The latter allows modification by electric fields and the study of thermalized bosonic quasiparticles, from the single particle level to interacting degenerate dense ensembles. Additionally, the freedom to combine bilayers of different van der Waals materials without lattice or relative twist angle constraints leads to layer hybridized and moir{\'e} excitons which can be widely engineered. This review covers fundamental aspects of IXs including correlation phenomena as well as the consequence of moiré superlattices with a strong focus on  TMD homo- and hetero-bilayers.
\end{abstract}


\maketitle
		
\section{Introduction} 
Van der Waals (vdW) bilayers host a plethora of different exciton species with different properties and functionalities that are promising for applications. Vice versa, excitons can serve as sensitive probes for the local potential landscape in vdW hetero-structures. Such vdW stacks can be prepared from the same or different two-dimensional crystals and are called homo-bilayer or hetero-bilayer, respectively. As Coulomb-bound electron-hole pairs, excitons are very sensitive to their dielectric environment inside and outside of the semiconducting host materials. Moreover, the excitonic properties strongly depend on the underlying Bloch states. Like in two-dimensional monolayers, a reduced dielectric screening and the two-dimensional nature of the bilayers results in excitonic binding energies exceeding several hundreds of millielectron volts. Unlike monolayers, vertical vdW double layers can host excitons with the constituting electron and hole states being localized in adjacent layers. As a direct consequence, the reduced overlap of the electron – hole wavefunctions results in enhanced lifetimes, up to several 100s of nanoseconds \cite{rivera_observation_2015, miller_long-lived_2017}, allowing the study of thermalized ensembles \cite{sigl_signatures_2020} of such interlayer excitons (IXs). 
Another peculiarity of IXs is their permanent dipole moment, which enables one to tune their energy and to control them by electric fields in capacitor structures \cite{ciarrocchi2019polarization} that can be realized even in devices prepared only from vdW materials with e.g. hBN as a dielectric material and graphene or thin graphite as semi-transparent electrodes. In addition to manipulating the IX energy, momentum-dependent hybridization of the IXs with other exciton species (in particular intralayer excitons with strong oscillator strengths) can be controlled by external stimuli such as electric fields \cite{kiemle2020control, 2023_Tagarelli_hybrid_IX_transport}, pressure \cite{brotons2018optical, 2023_Steeger_pressure_MoS2-BL}, or twist angle \cite{2023_Villafane_twist-MoSe2}. In this way, the competition and coupling between intralayer, interlayer and layer-hybridized excitons emitting at the fundamental gap as well as from higher lying states can be experimentally accessed \cite{2024_saigal_electric-controlIX}. While also lateral vdW hetero-structures can host IXs \cite{2018_Lau_lateral-IX}, this review focuses on the peculiarities of excitons in vertical vdW bilayers prepared from semiconducting transition metal dichalcogenide (TMD) monolayers. 
With several engineering opportunities, TMD-based vdW stacks hold fascinating properties that individual layers or conventional 3D solids typically do not exhibit within one material: (i) depending on the material combination, rich IX physics can be addressed including the dynamics of neutral, charged, and layer-hybridized IXs, as well as layer-specific characteristics of higher-lying IXs. (ii) Stacking of two lattices with similar but not identical lattice constants and/or a relative twist angle results in the formation of a moir{\'e} lattice. A moir{\'e} lattice is a geometrical superimposed periodic patterns that forms by stacking two atomically thin commensurate with a relative twist angle, or two crystals with different lattice constants or both together. The long-period moir{\'e} super-lattice structure causes highly periodic potential modulations in real space \cite{kang2013electronic} and flat moir{\'e} minibands in the reciprocal space for specific lattice mismatches and twist angles \cite{2018_Wu_moiré-band_theory}. The zero-dimensional potential traps can host so-called moir{\'e} excitons that can function as single photon emitters allowing for the creation of periodic networks of quantum emitters \cite{baek2020highly, 2024_Soltero_network}. Similar to magic angle twisted bilayer graphene \cite{2019_Lu_magic-angle-graphene}, in twisted bilayer TMDs, strongly correlated states can emerge by the combination of a quenched kinetic energy and a high density of states in the flat minibands, which can act as  sensitive probes for correlated phases of adjacent charge carrier systems \cite{2018_Wu_moiré-band_theory, 2020_Tang_Hubbard_WSe2-WS2, xu2020correlated, regan2020mott, 2023_Cai_FQAHE_trion-sensing_MoTe2}. Compared to intralayer excitons, IXs exhibit superior sensitivity due to their permanent dipolar moment \cite{liu2021excitonic}. In hetero-bilayers of TMDs, excitons can not only sense correlated states, but also correlated excitonic states, such as an excitonic insulator, can emerge \cite{2022_Chen_insulator, 2022_Zhang_corr_IX_ins, 2023_Xiong_corr-ins-phasediagram}.  
Moreover, the presence of strong spin-orbit coupling gives rise to  even more exotic correlated states
\cite{2021_Devakul_magic-tTMDs}. 
This high degree of tunability of rich exciton physics makes vertical vdW double layers a highly interesting research subject for fundamental studies on both many-body and correlation phenomena of low-dimensional exciton and charge carrier systems. On a more applied perspective, such highly tunable characteristics suggest various possibilities of optoelectronic and quantum photonic devices such as solar cells, light emitting diodes, photo-sensors or single-photon sources operating in a largely extended spectral range covering the near-infrared to ultraviolet wavelength ranges. This review article is organized as follows: After this introduction and a brief summary on the most fundamental properties of TMD semiconductors, IXs and dipolar excitons are discussed, followed by an introduction of excitons in TMD moir{\'e} hetero-structures and hybridization of intra- and interlayer excitons as well as the impact of structural effects. Next, interaction effects, including the formation of degenerate IX ensembles, are considered, followed by a discussion of IX formation and transport processes. The article concludes with a brief summary and an outlook. 

\section{Background}

Two-dimensional group-VI TMD semiconductors of the form MX$_2$ (with M = Mo, W, and X = S, Se) have attracted much  attention due to their appealing properties for a large palette of optoelectronics, spintronics, and photonics applications. The investigation of their structural, electronic, and optical properties constitutes a very active research field in the solid-state and photonics communities. Consequently, the main properties of 2D TMD semiconductors are extensively covered and reviewed in the literature for both their bulk and few layers forms \cite{yu2015valley,wang2018colloquium}. Among their main properties, TMDs monolayers are well known by their momentum-direct optical band gaps with energies in the visible to near-infrared spectral range, with the band edges located at the degenerate but inequivalent corners of the Brillouin zone (typically referred to as $\pm$K valleys). Carriers occupying the conduction and valence band edges at $\pm$K form excitons that are hydrogen-like states with a typical binding energy on the order of 0.5 eV \cite{chernikov2014exciton,he2014tightly}. Due to their large binding energy, excitons dominate the optical response of TMDs at both cryogenic and room temperatures. Moreover, the strong spin-orbit coupling induced by the heavy transition-metal atoms and the broken inversion symmetry of the TMDs lattice unit cell lead to an effective coupling between the valley index and spin of the electrons and holes at the $\pm$K corners \cite{xiao2012coupled}. This effective coupling (typically referred to as spin-valley locking), results in valley-dependent optical selection rules \cite{xu2014spin}: excitonic absorption and emission processes at $\pm$K involve $\sigma^{\pm}$-polarized photons, respectively, enabling optical control of excitons \cite{mak2012control,zeng2012valley}.  

\section{Interlayer excitons and dipoles}

\subsection{Interlayer excitons in natural multi-layer TMDs}

\begin{figure*}[t]
	\begin{center}
		\includegraphics[scale=0.45]{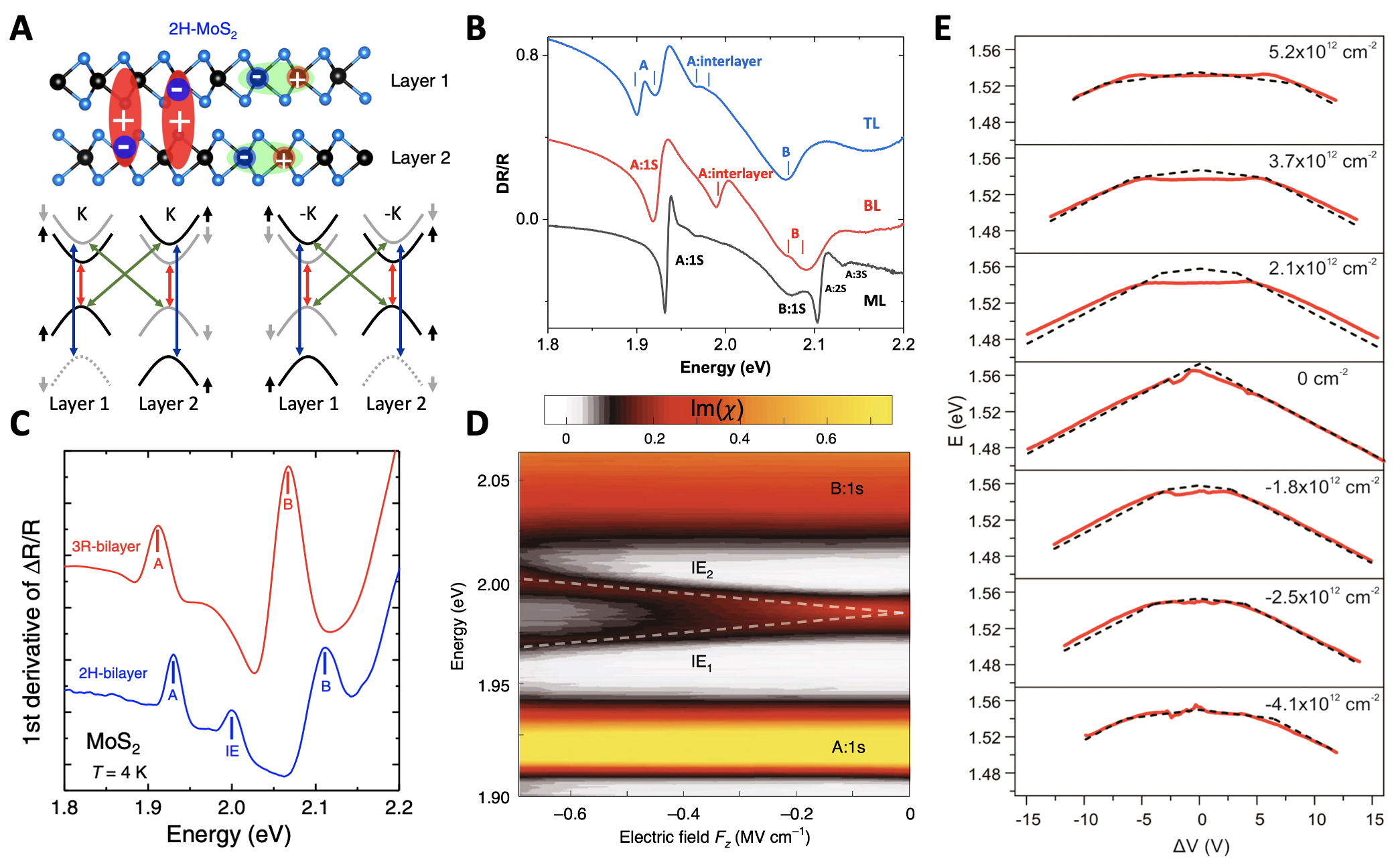}
	\end{center}
	\caption{Interlayer excitons in homobilayer TMDs [MoS$_2$ (A-D) and WSe$_2$ (E)].\\
	(A) Top: schematics showing the intralayer and interlayer nature of the excitons in bilayer 2H-MoS$_2$. Bottom: sketch of the spin, valley and layer configuration of the optical transitions corresponding to intralayer A excitons (red arrows), intralayer B excitons (blue arrows), and IXs (green arrows) in bilayer 2H-MoS$_2$. (B) Differential reflectivity of ML, BL and TL 2H-MoS$_2$ (black, red and blue solid lines, respectively) for a sample temperature of T = 4 K. (C) First derivative of the differential reflectivity spectrum for as-grown 2H-bilayer (blue) and as-grown 3R-bilayer (red) MoS$_2$ 4 K. (D) Color map of the absorption spectra of bilayer MoS$_2$ showing a Stark shift of the IXs at small electric fields. The intralayer A:1s and B:1s and the two branches of the IX resonances (IE1$_1$ and IE$_2$) are labelled. (E) IX emission energy in BL WSe$_2$ as a function of the applied electric field for different electron doping densities. Adapted with permission from Ref. \cite{paradisanos2020controlling}, Nature Publishing Group (A, C), Ref. \cite{gerber2019interlayer}, APS (B), Ref. \cite{leisgang2020giant}, Nature Publishing Group (B), and Ref. \cite{wang2018electrical}, ACS (E).}
	\label{Fig_3a}
\end{figure*}

Despite their indirect-gap nature,  multi-layer TMDs preserve the direct gap at the $\pm$K corners of the Brillioun zone even in the bulk limit, with spin-orbit-split conduction bands that present a flat dispersion along the out-of-plane direction (i.e., the K-H high-symmetry direction) \cite{arora2017interlayer,brotons2018optical}. Such flat band dispersion ensures that the electron wave functions around $\pm$K are highly confined within each individual layer, resulting in the formation of intralayer excitons, i.e. excitons in which the electron-hole pairs are localized in the same layer (see Figure \ref{Fig_3a}A). In addition to intralayer excitons, the layer-localized electron wave functions in TMD multi-layers can also bind to holes with wave functions confined within adjacent layers or that spread along several layers, giving rise to IXs with spatially displaced wave functions \cite{arora2017interlayer,wang2018electrical,horng2018observation,leisgang2020giant,lorchat2021excitons,peimyoo2021electrical}. Interlayer and intralayer excitons coexist in multi-layer TMDs (see Figure \ref{Fig_3a}A, top \cite{paradisanos2020controlling}), with both exciton species depicting momentum-direct optical transitions at $\pm$ K. The bottom panel of figure \ref{Fig_3a}A shows a schematic of the spin, band and layer configurations of dipole-allowed intralayer and IXs for bilayer 2$H$-MoS$_2$, a prototypical centrosymmetric multi-layer TMD. Ground-state intralayer excitons (the so called A excitons) show optical transitions involving an electron in the lowest conduction band and a hole in the highest valence band (red arrows), while intralayer B excitons present optical transitions between the topmost spin-orbit-split conduction band and the lower spin-orbit-split valence band at $\pm$K (blue arrows). Dipole allowed IXs present optical transitions between the topmost valence band in one layer and the upper spin-orbit-split conduction band in the other layer (green arrows). Consequently, dipole-allowed interlayer and intralayer excitons exhibit optical transitions with similar symmetry properties, with the difference that the spin-valley selectivity characteristic of monolayer TMDs is replaced by a spin-layer selectivity in the optical generation of excitons with circular polarized light. As detailed below, stacking-order dependent hybridization can give rise to a hole state that is delocalized across different layers \cite{paradisanos2020controlling, liang2022optically}.

Figure \ref{Fig_3a}B shows differential reflectance spectra (DR/R) of monolayer, bilayer and trilayer 2$H$-MoS$_2$ samples (black, red and blue solid lines, respectively) for a sample temperature of T = 4 K \cite{gerber2019interlayer}, in which an absorption peak arising from IXs can be seen for the bilayer and trilayer samples. Despite the large spatial separation of the electron and hole wavefunctions, IXs in multi-layer TMDs possess binding energies almost an order of magnitude larger compared with those in coupled III-V quantum wells, and of the same order of magnitude as their intralayer counterparts \cite{gerber2019interlayer}. Consequently, the optical transitions associated to IXs in TMDs can appear either at lower or at higher energies than the intralayer A exciton depending on the dark or bright nature of the ground IX state, respectively. Interestingly, the stacking order of the TMD multilayers has a strong impact on the formation of IXs. In Ref. \cite{paradisanos2020controlling}, the authors demonstrated that hole delocalization over MoS$_2$ homo-bilayers is allowed for 2$H$ stacking but forbidden for 3$R$ stacking, resulting in the presence (absence) of IXs in 2$H$-MoS$_2$ (3$R$-MoS$_2$) homo-bilayers (see Figure \ref{Fig_3a}C \cite{paradisanos2020controlling}). However, a more recent work showed the presence of IXs in $R$-MoS$_2$ bilayers depicting an  asymmetric interlayer coupling arising from a layer-dependent Berry phase effect. Such an asymmetric coupling is also the electronic origin of spontaneous polarization characteristic of R-stacked TMDs \cite{liang2022optically}.

Moreover, the spatial displacement of the exciton carriers in TMD homo-bilayers endows IXs with a large out-of-plane static electric dipole that can be tuned via the Stark effect \cite{lorchat2021excitons,leisgang2020giant,peimyoo2021electrical}. Figure \ref{Fig_3a}D depicts a color map of the absorption spectra of bilayer 2$H$-MoS$_2$ showing a Stark shift of the IXs for applied vertical electric fields \cite{leisgang2020giant}. Here, the IX resonance splits into two branches (IE$_1$ and IE$_2$) that shift linearly to lower (higher) energy depending on the parallel (anti-parallel) orientation of the electric field and the permanent electric dipole. The effects of an applied vertical electric field on the IX resonance energy and its recombination dynamics have also been investigated in dual-gated homo-bilayer WSe$_2$ and MoSe$_2$ devices \cite{wang2018electrical,feng2022highly}. Via photoluminescence (PL) and time-resolved PL spectroscopy at 10 K, the authors in Ref. \cite{wang2018electrical} show that the IX emission redshifts symmetrically for both positive and negative applied vertical electric fields. The symmetrical redshift of the exciton energy for both positive and negative fields is attributed to the electric-field-induced redistribution of carriers among the two layers that results in a parallel alignment of the exciton permanent dipole with the applied field. Figure \ref{Fig_3a}E summarizes the Stark shift observed under different doping densities (positive values \textit{n} for electron doping and negative values for hole doping) \cite{wang2018electrical}. Close to the charge neutrality point, a linear Stark effect emerges immediately, whereas for finite electron or hole doping densities an appreciable Stark effect is observed only for applied gate voltages beyond a threshold value. The existence of a threshold gate voltage can be understood as an offset in the resulting electric field produced by an unequal distribution of carriers between the two layers as a consequence of the applied field \cite{wang2018electrical}. 
Beyond natural bilayer TMD systems, ground and excited-state IXs with large permanent dipole moments were also been reported in trilayer 2H-MoSe$_2$ and three-, four-, five-, and seven-layer 2H-WSe$_2$, in which the wavefunctions of the carriers forming the excitons are confined in an every-other-layer configuration \cite{feng2022highly,zhang2023every}.
Finally, in addition to the Stark effect, the application of vertical electric fields can result in an enhancement of the IX recombination lifetime by more than two orders of magnitude \cite{wang2018electrical}.

\subsection{Excitons in TMD moir{\'e} hetero-structures}

In contrast to TMD homo-bilayers, stacking any two different ML TMDs creates a hetero-bilayer typically with type-II band alignment while preserving atomically sharp interfaces \cite{kang2013band,chiu2015determination,wilson2017determination}. Such a band alignment results in the formation of IXs with smaller transition energies than the intralayer excitons in the individual monolayers, from which the IXs localized at the $\pm$K valleys inherit their valley-contrasting physics  \cite{rivera2015observation,rivera2016valley,hanbicki2018double,ciarrocchi2019polarization}, although indirect IXs with an inter-valley nature also arise in many combinations of homo- and hetero-bilayers \cite{miller2017long}. In addition, the interlayer nature of the excitons leads to a reduced overlap of the electron and hole wave functions, which results in optical dipole transitions with long radiative lifetimes compared to intralayer excitons \cite{rivera2015observation,Nagler_2017,miller2017long}. Similar to IXs in TMD homo-bilayers, the separation of the exciton carriers can result in a large permanent electric out-of-plane dipole moment that enables a large tunability of the exciton transition energy by vertically applied electric fields \cite{ciarrocchi2019polarization,jauregui2019electrical,baek2020highly}. Moreover, similar to monolayer TMDs \cite{branny2017deterministic, palacios2017large}, patterned substrates were shown to lead to local strain profiles that enable exciton trapping in nanoscale confinement potentials, in which the mean number of trapped IXs can be controlled via the optical excitation level \cite{kremser2020discrete,li2020dipolar}. 

\subsubsection{Single interlayer exciton moir{\'e} trapping}

\begin{figure*}[!ht]
	\begin{center}
		\includegraphics[scale=0.58]{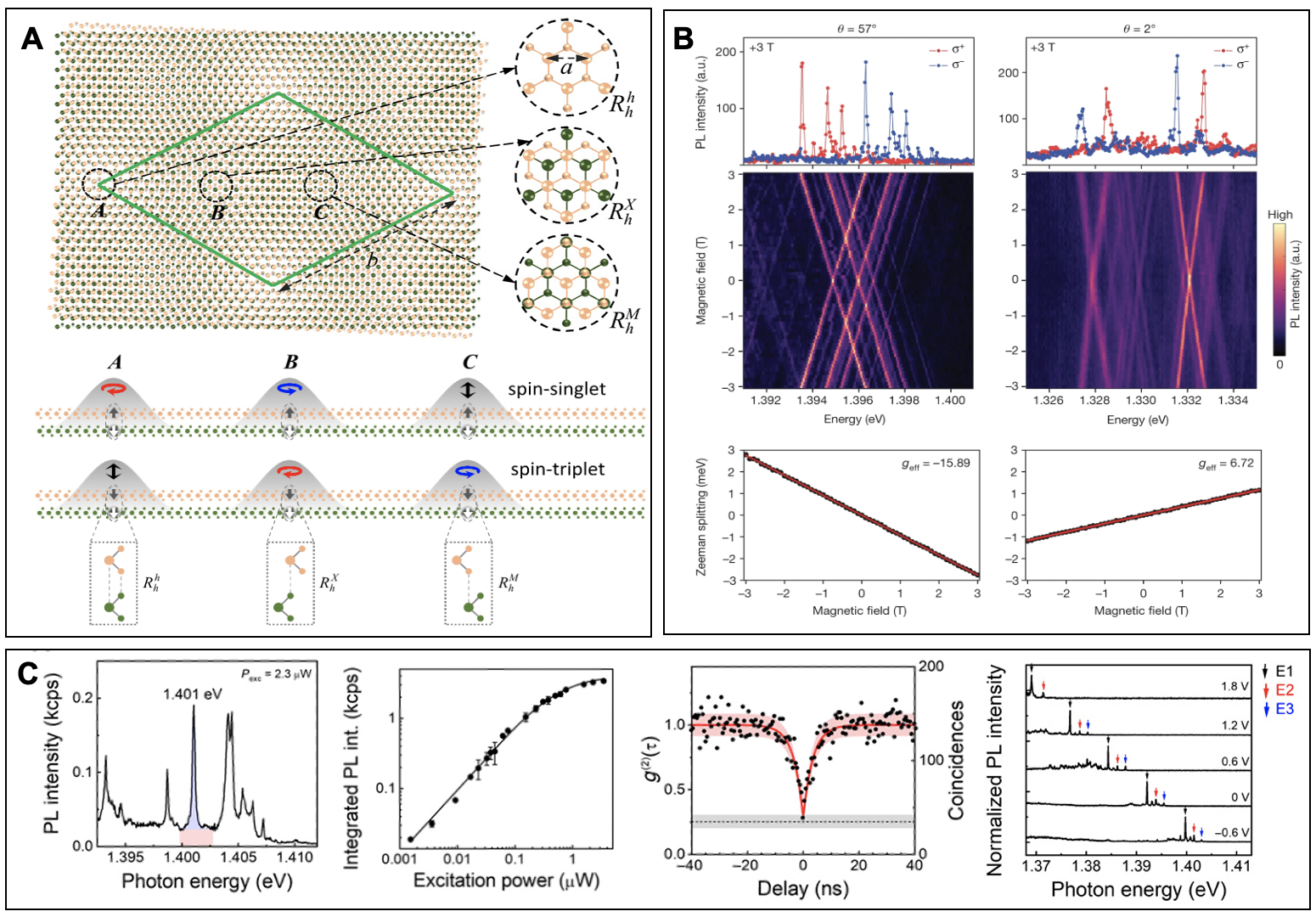}
	\end{center}
	\caption{Moir{\'e}-trapped IXs in MoSe$_2$/WSe$_2$ hetero-bilayers [MX$_2$/WX$_2$ (A) and MoSe$_2$/WSe$_2$ (B, C)].\\
	(A) Top: schematics of the long-period moir{\'e} pattern in a R-stacked MoX$_2$/WX$_2$ hetero-bilayer. Green diamond denotes the moir{\'e} supercell. Insets show zooms of the local atomic registries in the three high-symmetry trapping sites (A, B and C) of the moir{\'e} supercell. A, B and C have atomic registries $R^h_h$, $R^X_h$ and $R^M_h$, respectively. Bottom: schematic illustration of the polarization selection rules, for K-valley spin-singlet and spin-triplet IX wavepackets centered at trapping sites A, B and C, respectively. (B) Helicity-resolved PL of moir{\'e}-trapped IXs in MoSe$_2$/WSe$_2$ hetero-bilayers with twist angles of 57$^{\circ}$ (left) and 2$^{\circ}$ (right) as a function of the applied out-of-plane magnetic field (top panels). The bottom panels show the Zeeman splitting ($\Delta=E_{\sigma+}-E_{\sigma-}$) from the corresponding IXs as a function of the applied magnetic field with a linear regression from which the values of the effective Land{\'e} $g_{eff}$ of -15.89 and 6.72 are extracted. (C) Leftmost panel: PL spectrum showing a few moir{\'e}-trapped IXs in MoSe$_2$/WSe$_2$. The blue and red regions shown for the PL line at 1.401 eV represent the estimated PL signal from the emitter and the background, respectively. Second panel from the left: integrated PL intensity of the emitter highlighted in the leftmost panel at different excitation powers. Second panel from the right: second-order photon correlation statistics of the emitter highlighted in the leftmost panel. The red shadowed area represents the Poissonian interval error associated to the experimental determination of $g^{(2)}(\tau)$. The black dashed line and the gray shadowed area represent the average and error interval of the experimental limitation for $g^{(2)}(0)$, respectively, owing to the non-filtered emission background. Rightmost panel: PL spectra of moir{\'e}-trapped IXs in a dual-gated MoSe$_2$/WSe$_2$ hetero-bilayer at different gate voltages. Three representative peaks are indicated as E1, E2, and E3.}
	\label{Fig_3b}
\end{figure*}

Beyond the large permanent dipole moment and the spin-valley physics, the compelling concept of a moir{\'e} superlattice emerges in TMD hetero-bilayers with lattice mismatch and/or any relative twist angle between the constituent monolayers \cite{kang2013electronic}. The moir{\'e} superlattice creates a periodic potential landscape for IXs \cite{yu2017moire,wu2018theory,zhang2017interlayer}, in which three high-symmetry sites (A, B, and C) with specific atomic registries arise, such as R$_h^h$ (A), R$_h^X$ (B) and R$_h^M$ (C), where R$_h^\mu$ denotes an R-type stacking with the $\mu$ site of the electron layer (either $h$ the hexagon centre, $X$ the chalcogen site or $M$ the metal site) vertically aligned with the hexagon centre ($h$) of the hole layer (Figure \ref{Fig_3b}A). For moir{\'e} periods larger than the IX Bohr radius that is in the order of a few nanometers \cite{Deilmann_2020}, these moir{\'e} high-symmetry sites can behave as smooth quantum-dot-like confining potentials, leading to the trapping of single electrons, holes, or IXs \cite{yu2017moire,wu2018theory}. The moir{\'e} potential minima preserve the three-fold rotational ($C_3$) symmetry \cite{yu2017moire,wu2018theory}, and excitons trapped in such moir{\'e} sites obey selection rules that depend on the spin configuration of the exciton carriers (spin-singlet/spin-triplet) and the atomic registry of the trapping site, as theoretically predicted (see Figure \ref{Fig_3b}A) \cite{yu2018brightened}. Experimental evidence of neutral IXs trapped in a moir{\'e} potential has been reported in MoSe$_2$/WSe$_2$ hetero-bilayers with twist angles of around 0$^{\circ}$, 21.8$^{\circ}$ and 60$^{\circ}$ at cryogenic temperatures  \cite{seyler2019signatures,brotons2020spin,baek2020highly,barre2022optical}. Hetero-bilayers with twist angles of around 0$^{\circ}$ present ground IX states with spin-singlet configurations, while hetero-bilayers with twist angles of 21.8$^{\circ}$ and 60$^{\circ}$ present ground exciton states with spin-triplet optical transitions. For small IX densities, polarization-resolved PL measurements show that the moir{\'e}-trapped IXs exhibit linewidths below 100 $\mu$eV with strong helical polarization due to the $C_3$ symmetry, which results in a notable absence of observable fine structure \cite{seyler2019signatures,brotons2020spin,baek2020highly} (Figure \ref{Fig_3b}B, top). Moreover, the trapped IXs show well-defined magneto-optical properties: The g factors of the trapped excitons depend on the relative valley alignment (i.e., stacking configuration) between the layers hosting the carriers (Figure \ref{Fig_3b}B, bottom) \cite{seyler2019signatures,brotons2020spin,baek2020highly}. In addition, the emission from the localized interlayer excitons presents clear hallmarks of quantum-confined excitons: power-dependent emission intensities that can be described by a two-level saturation model \cite{seyler2019signatures,brotons2020spin,baek2020highly} and photon antibunching (Figure \ref{Fig_3b}C) \cite{baek2020highly}. Finally, the large permanent dipole of the moir{\'e}-trapped IXs can be exploited to achieve large direct current Stark tuning of their emission energies up to 40 meV \cite{baek2020highly}.

\subsubsection{Moir{\'e} interlayer trions}

\begin{figure*}[!ht]
	\begin{center}
		\includegraphics[scale=0.48]{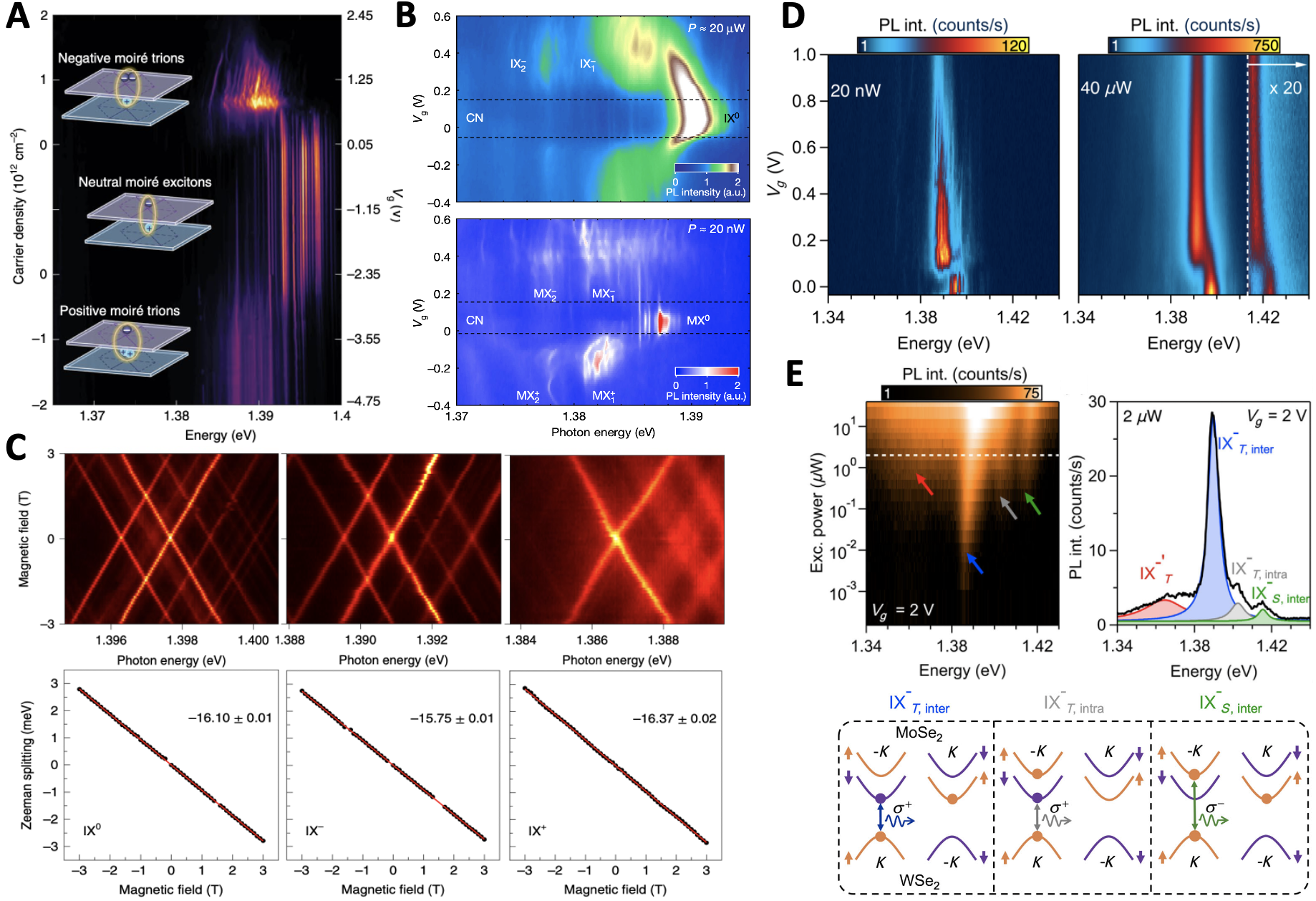}
	\end{center}
	\caption{Moir{\'e}-trapped interlayer trions in TMD hetero-bilayers [WSe$_2$/MoSe$_2$ (A-E)].\\
	(A) Gate-dependent PL emission from a WSe$_2$/MoSe$_2$ hetero-bilayer showing PL from moir{\'e}-trapped interlayer excitons and trions. The positive (negative) gate voltage ($V_g$) and carrier density  correspond to the electron (hole) doping. (B) Gate-dependent PL intensity of moir{\'e}-trapped IXs in a WSe$_2$/MoSe$_2$ hetero-bilayer for excitation powers of 20 $\mu$W (top) and 20 nW (bottom). (C) Top: PL spectra of moir{\'e}-trapped neutral (left), negatively-charged (middle), and positively-charged (right) IXs as a function of a vertical magnetic field. Bottom: Zeeman splitting of representative IXs extracted from the corresponding top panels. Land{\'e} $g_{eff}$ of -16.10, -15.75 and -16.37 are extracted for neutral IX$^0$, negatively charged IX$^-$ and positively charged IX$^+$, respectively. (D) Gate-voltage-controlled PL of IXs in the neutral (V$_g$ $\leq$ 0.1 V) and electron-doping regime (V$_g$ $\geq$ 0.1 V) for excitation powers of 20 nW (left panel) and 40 $\mu$W (right panel) at 4 K. In the right panel, the PL intensity is multiplied by a factor 20 in the spectral range delimited by the vertical dashed area for visualization purposes. (E) Right: Logarithmic-scale density plot showing the  evolution of the PL spectrum of negative interlayer trions as a function of excitation power, in which four different peaks can be resolved (as indicated by the arrows). Left: PL spectrum acquired for an intermediate excitation power of 2 $\mu$W (as indicated by the white dashed line in the left panel) showing four Lorentzian peaks corresponding to various exciton species. Bottom: Schematic representation of the charge configurations for IX$^-_{T,inter}$ (left), IX$^-_{T,intra}$ (middle), and IX$^-_{S,inter}$ (right) showing the optical transitions that involve a hole in the topmost valence band of WSe$_2$ at $K$. Adapted with permission from Ref. \cite{Wang.2021} (A), Ref. \cite{liu2021signatures} (B), Ref. \cite{baek2021optical} (C), Nature Publishing Group, and Ref. \cite{brotons2021moir} (D, E), American Physical Society.}
	\label{Fig_3c}
\end{figure*}

In addition to neutral IXs, the loading of additional charge carriers into the moir{\'e} lattice in gate-tunable WSe$_2$/MoSe$_2$ hetero-bilayers also enables the formation of moir{\'e}-trapped interlayer trions, which are bound quasiparticles consisting of two electrons and a hole (negatively charged trion) or a single electron and two holes (positively charged trion)  \cite{liu2021signatures,brotons2021moir,Wang.2021,baek2021optical}. Upon electron or hole doping of the hetero-bilayer, the neutral trapped IXs form on-site negatively or positively charged interlayer trions with an average binding energy of $\sim$7 and $\sim$6 meV, respectively (see Figure \ref{Fig_3c}A) \cite{liu2021signatures,brotons2021moir,Wang.2021,baek2021optical}. Interestingly, the doping-dependent evolution of the IX PL shows the same overall behavior for both low and high IX densities (Figure \ref{Fig_3c}B) \cite{liu2021signatures,brotons2021moir}, with the only differences being: i) the emission line widths, which are one order of magnitude broader for the ensemble exciton peaks; and ii) the absolute emission energies, which show an exciton-density-dependent blue-shift in the ensemble exciton peaks as a consequence of the dipolar interactions \cite{brotons2021moir}.
Moreover, the magneto-optical properties of the neutral and charged IXs are also identical (see Figure \ref{Fig_3c}C). However, at high IX densities, emission from a higher energy IX ensemble peak with opposite selection rules and different effective g-factor appears, which has been attributed to optical transitions involving the highest spin-orbit-split conduction band of MoSe$_2$ at $\pm$K \cite{ciarrocchi2019polarization,wang2019giant,joe2021electrically,brotons2021moir}. This higher energy ensemble exciton state can also be tuned from a neutral to a charged regime by gate doping, with a charging offset very similar to the ground IX state at lower energy (see Figure \ref{Fig_3c}D)\cite{brotons2021moir}.

Finally, a combination of power-dependent and magneto-optical measurements in a gate-tunable MoSe$_2$/WSe$_2$ hetero-bilayer revealed the presence of three different species of moir{\'e}-localized negative trions with contrasting spin-valley configurations: intervalley interlayer trions with spin-singlet optical transitions, and both intervalley and intravalley interlayer trions with spin-triplet optical transitions (as schematically depicted in the right, central, and left panels of the bottom of Figure \ref{Fig_3c}E, respectively \cite{brotons2021moir}), which result in PL spectra that can show up to four different PL components: the three moir{\'e}-localized negative trions with contrasting spin-valley configurations plus an additional low-energy peak which has recently been attributed to a composite six-particle “hexciton” state \cite{van2022six}.

\subsubsection{Moir{\'e} effects on the intralayer excitons}

\begin{figure*}[t]
	\begin{center}
		\includegraphics[scale=0.45]{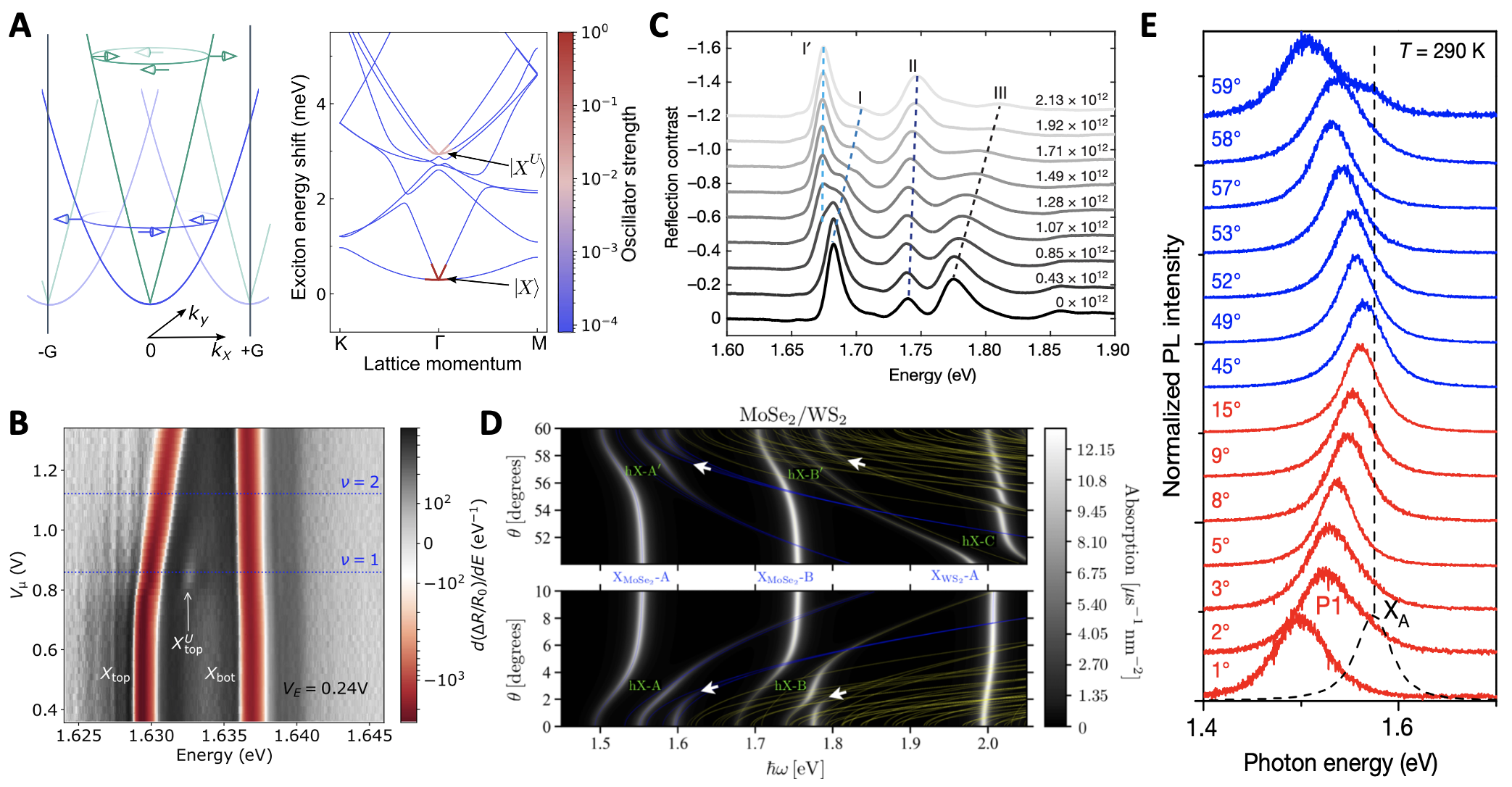}
	\end{center}
	\caption{Effects of the moir{\'e} superlattice on the intralayer excitons [MoSe$_2$/hBN/MoSe$_2$ (B), WS$_2$/WSe$_2$ (C), MoSe$_2$/WS$_2$ (D,E)].\\
	(A) Left: bare dispersion of mobile excitons in a ML TMD. The linearly polarized exciton modes split into two branches with linear (green line) and parabolic (blue line) dispersion due to the strong intervalley electron-hole exchange coupling. The transparent lines represent higher bands arising from the mixing of states connected by reciprocal lattice vectors as a consequence of a periodic potential. Right: dispersion bands of excitons moving in periodic moir{\'e} potential along a path in the moir{\'e} Brillouin zone. The color bar indicates the oscillator strength of each state (saturated for all blue lines). Only a single Umklapp band per polarization obtains sizable oscillator strength, while most states remain dark. (B) Dependence of the differential reflectance measured in MoSe$_2$/hBN/MoSe$_2$ hetero-structure as function of the chemical potential $V_{\mu}$ differentiated with respect to energy $E$ for a fixed vertical electric field $V_E=$ 0.24 V. An additional higher-energy Umklapp exciton resonance $X^U_{top}$ is observed once the top layer is filled with one electron per moir{\'e} site (i.e., $\nu=1$). (C) Reflection contrast spectrum measured in a moir{\'e} WS$_2$/WSe$_2$ hetero-bilayer as a function of electron concentration. Three prominent moir{\'e} exciton peaks are labelled. Peak I exhibits a strong blue-shift and diminishes upon doping while another  lower energy peak (I${'}$) emerges. Similarly peak III shows a strong blue-shift and weakens upon doping. (D) Calculated absorption spectrum as functions of twist angle for MoSe$_2$/WS$_2$ close to parallel (bottom panel) and anti-parallel (top panel) alignment. The full low-energy exciton spectrum is overlaid on top the absorption map by blue and yellow curves, showing multiple momentum-dark exciton states. (E) Normalized room-temperature PL spectra of MoSe$_2$/WS$_2$ hetero-bilayers with interlayer twist angles ranging from 1$^{\circ}$ to 59$^{\circ}$. The dashed curve shows the typical room-temperature PL peak from the A exciton of ML MoSe$_2$ (X$_A$), and its energy is indicated by the vertical dashed line. Adapted with permission from Ref. \cite{shimazaki2021optical} (A, B), Ref. \cite{jin2019observation} (C), Ref. \cite{Tijerina2019} (D), Ref. \cite{alexeev2019resonantly} (E).}
	\label{Fig_3d}
\end{figure*}

The presence of a long-range periodic exciton potential can profoundly modify the optical spectrum of TMDs. This is the case for twisted TMD homo- and hetero-bilayers, in which the moir{\'e} superlattice gives rise to a long-range crystal structure with a reduced Brillouin zone. The reduced Brillouin zone results in a folding-induced flattening of the conduction and valence bands and a multitude of avoided crossings that arise as a consequence of the interlayer hybridization \cite{wu2017topological,jin2019observation}. Recently, several groups have experimentally shown that the periodic moir{\'e} exciton potential leads to a mixing of momentum states separated by moir{\'e} reciprocal lattice vectors, which results in the formation of satellite exciton peaks (see Figure \ref{Fig_3d}A) \cite{jin2019observation,shimazaki2021optical}. Although the moir{\'e} superlattice plays a key role in the formation of the satellite peaks observed in these works (also referred to as Umklapp \cite{shimazaki2021optical} or moir{\'e} excitons \cite{jin2019observation}), the origin of the periodic exciton potential is different. In Ref. \cite{shimazaki2021optical}, the authors used a twisted MoSe$_2$/hBN/MoSe$_2$ homo-bilayer structure where the carrier density in the two MoSe$_2$ layers can be controlled independently. The presence of the monolayer-thick hBN barrier layer in their device reduces the interlayer coupling between the twisted MoSe$_2$ layers, which results in a weak periodic moir{\'e} potential for intralayer excitons. Therefore, in the absence of electron or hole doping, the spectrum does not show any moir{\'e} exciton peaks (see Figure \ref{Fig_3d}B). However, for unity electron filling of the underlying moir{\'e} potential in either or both MoSe$_2$ layers, new optical resonances appear in the reflection spectrum (see Figure \ref{Fig_3d}B). Such Umklapp or moir{\'e} exciton resonances arise due to the spatially modulated interactions between excitons and electrons in an incompressible Mott-like correlated state, which creates a periodic potential for excitons with the periodicity imposed by the moir{\'e} lattice constant \cite{shimazaki2021optical}. In contrast to the MoSe$_2$/hBN/MoSe$_2$ device in Ref. \cite{shimazaki2021optical}, nearly-aligned WSe$_2$/WS$_2$ hetero-bilayers present a strong static moir{\'e} potential, which leads to the formation of satellite moir{\'e} exciton peaks even in the absence of electron or hole doping (see Figure \ref{Fig_3d}C) \cite{jin2019observation}. The presence of such satellite moir{\'e} exciton peaks is not restricted to hetero-structures consisting of two layers, but has also been reported in devices in which the WSe$_2$ layer is replaced by a bilayer and a trilayer \cite{chen2022tuning}.

Moreover, the moir{\'e} pattern of TMD hetero-bilayers has shown to result in the formation of non-trivial many-body excitonic states \cite{naik2022intralayer,wang2023intercell}. Examples of such many-body moir{\'e} excitons include intralayer charge-transfer excitons \cite{naik2022intralayer} and an interlayer moir{\'e} exciton in which the hole's wavefunction is surrounded by the corresponding electron’s wavefunction, which is distributed among three adjacent moir{\'e} traps \cite{wang2023intercell}.

\begin{figure*}[t]
	\begin{center}
		\includegraphics[scale=0.4]{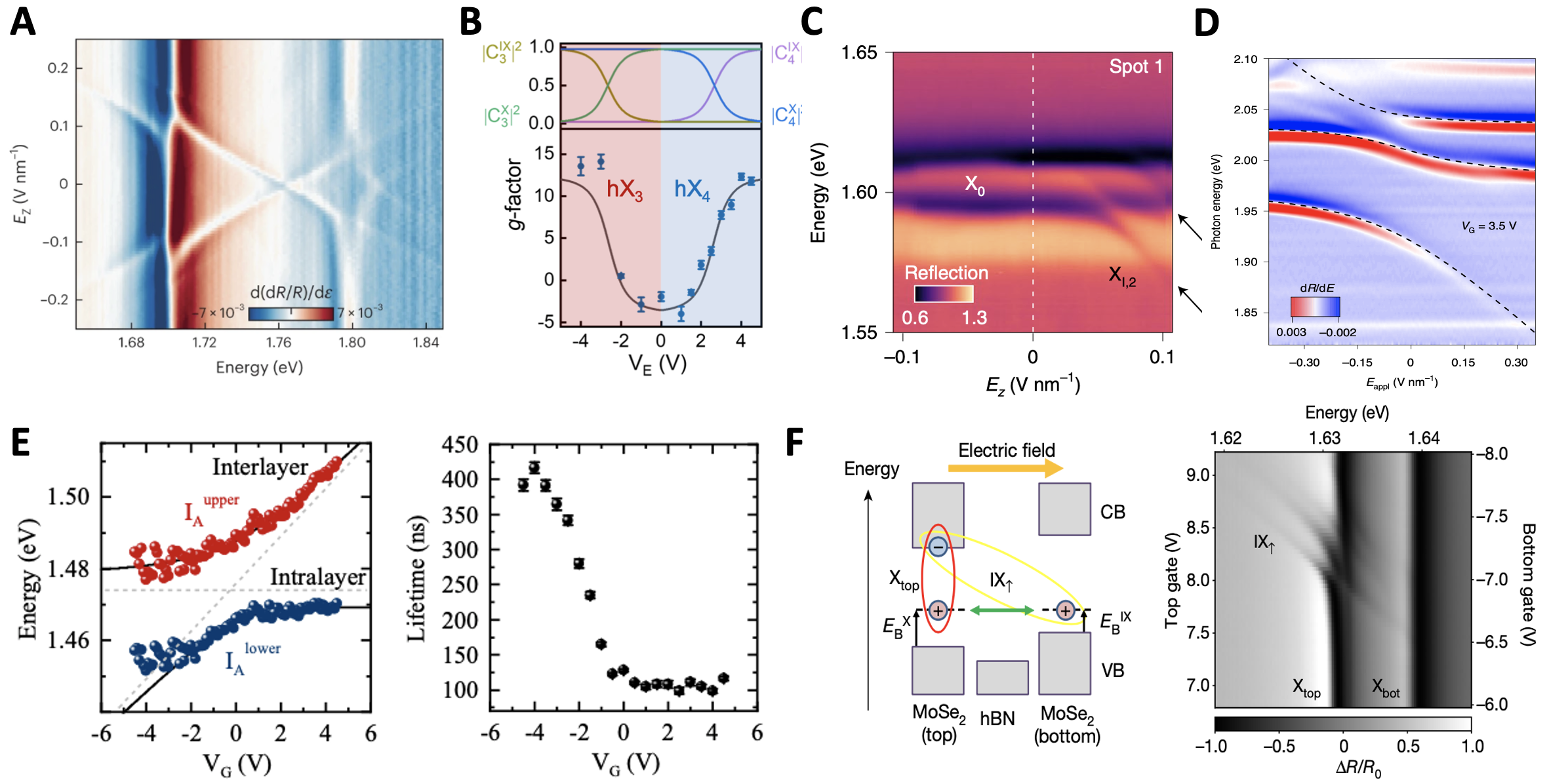}
	\end{center}
	\caption{Hybridization of intralayer and interlayer excitons [3L 2H-WSe$_2$ (A), 2L 2H-MoSe$_2$ (B), twisted (t$\approx 0^{\circ}$) MoSe$_2$/MoSe$_2$  (C), twisted (t$\approx 60^{\circ}$) WSe$_2$/WS$_2$ (D), MoS$_2$/WS$_2$ (E), MoSe$_2$/hBN/MoSe$_2$ (F)].\\
	(A) Color maps of the reflectance spectrum differentiated with respect to photon energy from trilayer 2H-WSe$_2$ as a function of the electric field $E_z$ applied perpendicular to the trilayer in the spectral regions around the ground and first-excited state of the A-exciton. (B) Electric-field-driven evolution of the g-factor of the hybrid inter- and intralayer excitons hX$_{3}$ and hX$_{4}$ in a natural 2H-MoSe$_2$ bilayer (bottom panel). The top panel shows the electric-field-dependent normalized contributions of each bare exciton state $|C^{IX(X)}|^2$ to the corresponding hybrid excitons. (C) Electric-field-dependent reflectance contrast spectra of the intralayer excitons in a near-0$^{\circ}$-twist-angle MoSe$_2$/MoSe$_2$ bilayer which features large rhombohedral AB/BA domains. (D) Energy derivative of the reflectance contrast spectrum as a function of the applied out-of-plane electric field in a 60$^{\circ}$ aligned WSe$_2$/WS$_2$ sample at a fixed doping of one electron per moir{\'e} cell. (E) Left panel: Emission energy of the upper (I$_{A}^{upper}$) and lower exciton branches (I$_{A}^{lower}$)  in a MoS$_2$/WS$_2$ hetero-bilayer embedded in an vdW field-effect structure as a function of the applied gate voltage. Right panel: Hybridization-induced evolution of the PL lifetime corresponding to the lower exciton branch measured at a bath temperature of 10 K. (F) Color map of the reflectance signal from a MoSe$_2$/hBN/MoSe$_2$ as a function of the applied vertical electric field (right panel) together with the corresponding schematics of the energy bands and the exciton energy alignment (left panel). Adapted with permission from Ref. \cite{zhang2023every} (A), Ref. \cite{feng2022highly} (B), American Physical Society, Ref. \cite{sung2020broken} (C), Ref. \cite{tang2021tuning} (D), Ref. \cite{kiemle2020control} (E), and Ref. \cite{shimazaki2020strongly} (F), Nature Publishing Group.}
	\label{Fig_3e}
\end{figure*}

In addition to the satellite spectral features caused by Umklapp scattering within the reduced Brillouin zone and the presence on exotic many-body moir{\'e} excitons, the moir{\'e} effects on the intralayer excitons in TMD hetero-bilayers that feature a close energy alignment of the band edges (such as MoSe$_2$/WS$_2$ and MoTe$_2$/MoSe$_2$) are enhanced by resonant interlayer band hybridization \cite{Tijerina2019}. Theoretical calculations show that the resonantly hybridized exciton energy in these TMD hetero-bilayers shows a sharp modulation as a function of the interlayer twist angle (see Figure \ref{Fig_3d}D) \cite{Tijerina2019}. Such resonant hybridization between exciton bands has been experimentally observed through PL measurements in WS$_2$/MoSe$_2$ \cite{alexeev2019resonantly}, MoS$_2$/WSe$_2$ \cite{nagler2019interlayer} and WS$_2$/WSe$_2$ \cite{wu2021identification} hetero-bilayers, which feature nearly-degenerate band edges. The hybridized excitons in these systems exhibit a pronounced energy shift as a function of twist angle (see Figure \ref{Fig_3d}E) \cite{alexeev2019resonantly}.

\subsection{Hybridization of intralayer and interlayer exciton states}

Intralayer and interlayer excitons in multi-layer TMDs can hybridize even in the absence of a moir{\'e} superlattice. The large quantum-confined Stark effect of IXs in TMD homo- and hetero-bilayers allows one to exploit the application of vertical electric fields to energetically tune IXs into resonance with intralayer excitons, where they hybridize \cite{leisgang2020giant,sung2020broken,kiemle2020control,tang2021tuning,lorchat2021excitons,peimyoo2021electrical,sponfeldner2021capacitively,zhang2023every,feng2022highly,Polovnikov_2024}. In the case homo-bilayer MoS$_2$, the results in Ref. \cite{leisgang2020giant} show that the Stark-split IX branches undergo clear avoided crossings with both the intralayer B- and A-exciton branches, where the coupling orginates from hole tunnelling and from an exchange-induced A-B exciton admixture, respectively \cite{sponfeldner2021capacitively}. Similar couplings between interlayer and intralayer excitons have also been reported by other groups in bilayer 2H-MoS$_2$ \cite{lorchat2021excitons,peimyoo2021electrical}, natural bilayer and trilayer 2H-MoSe$_2$ \cite{feng2022highly}, and three-, four-, five-, and seven-layer natural 2H-WSe$_2$ \cite{zhang2023every}. Figure \ref{Fig_3e}A shows a density plot of the electric field dependence of the reflectance spectrum differentiated with respect to photon energy from trilayer 2H-WSe$_2$, where the hybridization of the IXs with the 1s and 2s states of the intralayer A exciton leads to clear anticrossings at the corresponding intralayer exciton energies \cite{zhang2023every}.

In addition to the normalization of the exciton resonance energy and the redistribution of oscillator strength between the different exciton branches, the hybridization of interlayer and intralayer excitons in natural multilayer 2H-stacked TMDs has also shown to result in gate-tunable $g$ factors for the resulting hybrid exciton species \cite{lorchat2021excitons,feng2022highly}. Figure \ref{Fig_3e}B shows the electric-field-driven evolution of the g-factor of the hybrid excitons in a natural 2H-MoSe$_2$ bilayer \cite{feng2022highly} (bottom panel). The top panel shows the electric-field-dependent normalized contributions of each bare exciton state $|C^{IX(X)}|^2$ to the corresponding hybrid excitons, showing a clear evolution from pure intralayer (interlayer) to pure interlayer (intralayer) character before and after the anticrossing.

The $V_g$-induced hybridization between intralayer and interlayer excitons is not restricted to natural TMD multilayers, but can also be observed in twisted TMD homo- and hetero-bilayers \cite{sung2020broken,kiemle2020control,tang2021tuning}. In Ref. \cite{sung2020broken}, the coupling between intralayer and interlayer excitons in near-0$^{\circ}$-twist-angle MoSe$_2$/MoSe$_2$ homo-bilayers is reported. These homo-bilayers featured large rhombohedral AB/BA domains, which support IXs with out-of-plane electric dipole moments in opposite directions that can be flipped by the application of vertical electric fields, resulting in field-asymmetric hybridization with intralayer excitons (see Figure \ref{Fig_3e}C). In Ref. \cite{tang2021tuning},  hybridization of interlayer and moir{\'e} excitons in angle-aligned WSe$_2$/WS$_2$ and MoSe$_2$/WS$_2$ hetero-bilayers was observed and in Ref. \cite{Polovnikov_2024} for MoSe$_2$/WS$_2$ hetero-bilayer. The hybrid excitons are formed via spin-conserving resonant tunnelling of electrons or holes between the layers, and exhibit the characteristics of both interlayer (large out-of-plane electric dipole) and intralayer excitons (appreciable oscillator strength). Figure \ref{Fig_3e}D shows the electric-field dependence of hybrid excitons in a WSe$_2$/WS$_2$ moir{\'e} superlattice loaded with one electron per site \cite{tang2021tuning}, where the energy-level anticrossing between the intralayer moir{\'e} excitons (higher energy peaks) and the IXs (lower energy peak) can be observed. Similar electric-field-induced hybridization between interlayer and intralayer excitons was observed in angle-aligned MoS$_2$/WS$_2$ hetero-bilayers embedded in a vdW field-effect structure (see Figure \ref{Fig_3e}E) \cite{kiemle2020control}. In addition to the electric-field-induced energy anti-crossing between the upper and lower exciton branches in the PL spectra, the authors in Ref. \cite{kiemle2020control}
also showed hybridization-induced renormalization of the PL lifetime for the lower exciton emission branch (see Figure \ref{Fig_3e}E). The electric-field-induced renormalization of the hybrid exciton resonance energies in all these systems can be quantitatively reproduced by a phenomenological model in which the hybridization between different exciton states is treated as a coupling between oscillators with resonance energies corresponding to the bare exciton states being hybridized \cite{kiemle2020control,lorchat2021excitons,tang2021tuning}. The black dashed lines in Figures \ref{Fig_3e}D and \ref{Fig_3e}E show the best fits of the experimental data to an oscillator model, from which inter-/intralayer couplings as high as 40 meV and 11 meV could be extracted, respectively, demonstrating that these systems were strongly coupled \cite{tang2021tuning,kiemle2020control}.

Finally, the coupling between interlayer and intralayer excitons has been observed even for hetero-structures in which the interlayer exciton constituent states are located in layers separated by a hBN tunnel barrier \cite{shimazaki2020strongly}. Using a double-gated MoSe$_2$/hBN/MoSe$_2$ hetero-structure, coherent coupling of interlayer and intralayer excitons via hole tunnelling through the hBN barrier has been explored. Figure \ref{Fig_3e}F shows a color map of the reflectance signal from the MoSe$_2$/hBN/MoSe$_2$ as a function of the applied vertical electric field, where the resonances at 1.632 eV and 1.640 eV correspond to intralayer excitons in the top and bottom MoSe$_2$ layers, respectively \cite{shimazaki2020strongly}. At large applied electric fields, several resonances with a strong E-field dependence are observed, which originate from IXs with a large dipole moment leading to a sizeable Stark shift. The spectra for a positive top gate voltage (V$_{tg}$) regime correspond to the IX$_{\uparrow}$, which have a hole in the bottom layer and an electron in the top layer (see the schematic of the energy bands and the exciton energy alignment under electric fields in Figure \ref{Fig_3e}F). Interestingly, the results in Figure \ref{Fig_3e}F show that IX$_{\uparrow}$ hybridize exclusively with intralayer excitons in the top layer, as seen by a multitude of avoided crossings, which unequivocally shows that the coupling originates exclusively from spin-conserving hole tunnelling. Finally, the  existence of multiple avoided crossings demonstrates the existence of a moir{\'e} superlattice in the MoSe$_2$/hBN/MoSe$_2$ hetero-structure.

\subsection{Structural effects on the PL of interlayer excitons}

\begin{figure*}[t]
	\begin{center}
		\includegraphics[scale=0.45]{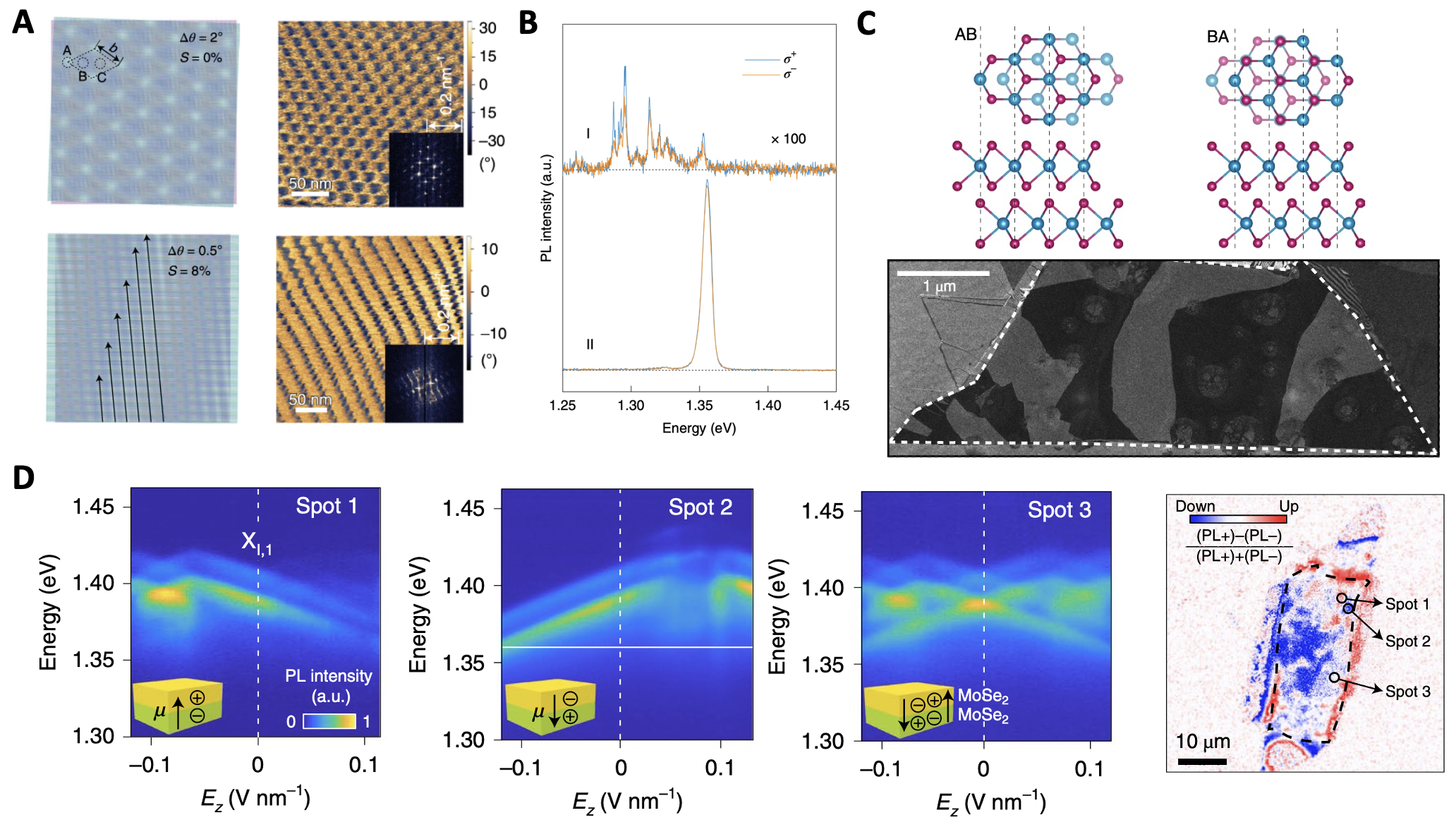}
	\end{center}
	\caption{Structural effects on the PL of interlayer excitons [MoSe$_2$/WSe$_2$ (A, B) and twisted MoSe$_2$/MoSe$_2$ (C, D)].\\
	(A) Top left: sketch of a MoSe$_2$/WSe$_2$ hetero-bilayer with a twist angle of 2$^{\circ}$ at no applied strain. The dashed lines indicate the supercell of the hexagonal moir{\'e} superlattice, where A, B and C represent high-symmetry points preserving the $C_3$ symmetry within the supercell. Bottom left: skecth of a MoSe$_2$/WSe$_2$ hetero-bilayer with a twist angle of 0.5$^{\circ}$ under an applied uniaxial hetero-strain of 8$\%$. The arrows indicate the resulting primary 1D moir{\'e} structures. Right panels: experimental piezoresponse force microscopy (PFM) images of the same spatial spot in a MoSe$_2$/WSe$_2$ hetero-bilayer with a twist angle of $\sim$59$^{\circ}$ under no applied (top) and applied uniaxial strain (bottom). (B) Representative PL spectra from two MoSe$_2$/WSe$_2$ hetero-bilayer samples with h-BN encapsulation showing circularly-polarized quantum emitter-like sharp PL peaks (top) and linearly-polarized ensemble IX emission (bottom). The blue and orange lines represent helicity-resolved spectra for $\sigma^+$ and $\sigma^-$ emission, respectively, under $\sigma^+$ excitation. (C) Top panels: top and side views of the atomic structure in a twisted MoSe$_2$/MoSe$_2$ homo-bilayer showing rhombohedral AB (left) and BA stacking configurations (right). Bottom panel: dark-field TEM image of a near-0$^\circ$-twist-angle MoSe$_2$/MoSe$_2$ bilayer showing alternating, micrometre-sized AB and BA domains. (D) Electric-field-dependent PL spectra of IXs measured at three different spots of a twisted MoSe$_2$/MoSe$_2$ homo-bilayer (first three panels from the left). The insets show sketches of the predominant orientation of the permanent electric dipole at each spot. The right panel shows a map of the magnitude (PL$^+$-PL$^-$)/(PL$^+$+PL$^-$), which is proportional to the orientation (up/down) of the permanent electric dipole. PL$^\pm$ is the PL intensity at $E_z$ = $\pm0.15$ V nm$^{-1}$, integrated over the energy range below 1.36 eV. Adapted with permission from Ref. \cite{bai2020excitons} (A, B), and Ref. \cite{sung2020broken} (C, D), Nature Publishing Group.}
	\label{Fig_3f}
\end{figure*}

Beyond the effects of applied external electric fields, carrier doping, layer- and exciton-hybridization, and the moir{\'e} superlattice discussed in the previous sections, the structural properties of the TMD hetero-structures such as strain and atomic reconstruction also play an important role on the optical and electric properties of their host IXs. Recently, it has been shown that the PL of IXs in TMD homo- and hetero-bilayers can be profoundly affected by the structural properties of the hetero-structure \cite{bai2020excitons,sung2020broken,2023_Zhao_mesoco-reconstr, 2023_Li_reconstruction-hoegele}. In Ref. \cite{bai2020excitons}, the authors used real-space imaging to show how the application of uniaxial hetero-strain in a WSe$_2$/MoSe$_2$ moir{\'e} hetero-bilayer, where the weak vdW interaction between the layers can result in different deformation, leads to a transition from a triangular moir{\'e} lattice of zero-dimensional traps into parallel stripes of one-dimensional quantum wires. Figure \ref{Fig_3f}A illustrates the concept and shows experimental results for a nominally unstrained (top panels) and a strained hetero-bilayer (bottom panel). Interestingly, optical spectroscopy characterization of the samples revealed that the IX PL changes drastically from the unstrained to the strained samples. Figure \ref{Fig_3f}B shows representative low-temperature PL spectra measured in nominally unstrained (top) and strained hetero-bilayers (bottom) for helicity-resolved photon collection. As can be seen in these plots, the PL emission from IXs in strained samples (i.e., depicting one-dimensional moir{\'e} potentials) shows linear polarization and two orders of magnitude higher intensity than the circularly-polarized quantum emitter-like sharp PL peaks characteristic of the zero-dimensional moir{\'e} traps.

In addition to active approaches to modify the IX emission through structural modifications of the TMD hetero-structures, intrinsic structural effects such as atomic reconstruction and domain formation in twisted hetero-structures \cite{enaldiev2020stacking,weston2020atomic,mcgilly2020visualization,rosenberger2020twist,2021_Andersen_moire_sem,2023_Zhao_recnstructed-eciton} can also strongly alter the IX properties. In Ref. \cite{sung2020broken}, a combination of electronic and optical far-field spectroscopy is employed to study near-0$^\circ$-twist-angle MoSe$_2$/MoSe$_2$ homo-bilayers featuring large rhombohedral AB/BA domains (see the sketch in the top panel of Figure \ref{Fig_3f}C). The bottom panel of Figure \ref{Fig_3f}C shows a dark-field TEM image of one of their near-0$^\circ$-twist-angle MoSe$_2$/MoSe$_2$ homo-bilayers, where alternating, micrometre-sized AB and BA domains can be observed due to the effects of atomic reconstruction. Further, the authors showed that the broken mirror/inversion symmetry exhibited by the alternating AB/BA domains results into an effective locking of the domain atomic stacking and the orientation of the IX permanent dipole moment: AB (BA) domains host IXs with permanent electric dipole pointing up (down) \cite{sung2020broken}. This effect leads to opposite energy Stark shifts for IXs in AB and BA domains under applied vertical electric fields. Figure \ref{Fig_3f}D shows electric-field-dependent PL spectra of IXs measured at three different spots in a twisted MoSe$_2$/MoSe$_2$ homo-bilayer (first three panels from the left) \cite{sung2020broken}. As can be seen in the first two panels, the IXs in two of the measured spots show field-dependent Stark shifts with similar magnitude but opposite signs, indicating that the excitons in the two spots presented opposite dipole moments (as schematically shown in the insets). Interestingly, these measurements also showed that the dipole orientation in both spots can be flipped at large enough applied fields. The results of a third spot demonstrate IX with Stark shifts with both positive (blue-shift) and negative (red-shift) slopes, indicating the presence of both AB and BA domains inside the dimensions of the confocal PL spot. The authors were also able to exploit the domain-dependent sign of the IX Stark shift to generate a spatial map of the electric dipole moment of the IX in their sample (see right panel in Figure \ref{Fig_3f}D). 

Finally, in Ref. \cite{2023_Zhao_recnstructed-eciton} the authors showed one-to-one correlations between local spectral features and sample morphology in non-gated MoSe$_2$/WSe$_2$ hetero-bilayers, which suggests the co-existence of domains of different dimensionality and exciton characteristics. Their results show that reconstructed 2D domains with large lateral dimensions in small-twist hetero-bilayers exhibit clear luminescent singlet and triplet IXs while split intralayer exciton resonances and spectrally narrow IXs are present in the 1D domains connecting the extended 2D domains to arrays of nanometre-sized 0D domains.

\section{Interactions}

\subsection{Interlayer excitons interacting with electronic states in moir{\'e} hetero-structures}

\begin{figure*}[t]
	\begin{center}
		\includegraphics[scale=0.43]{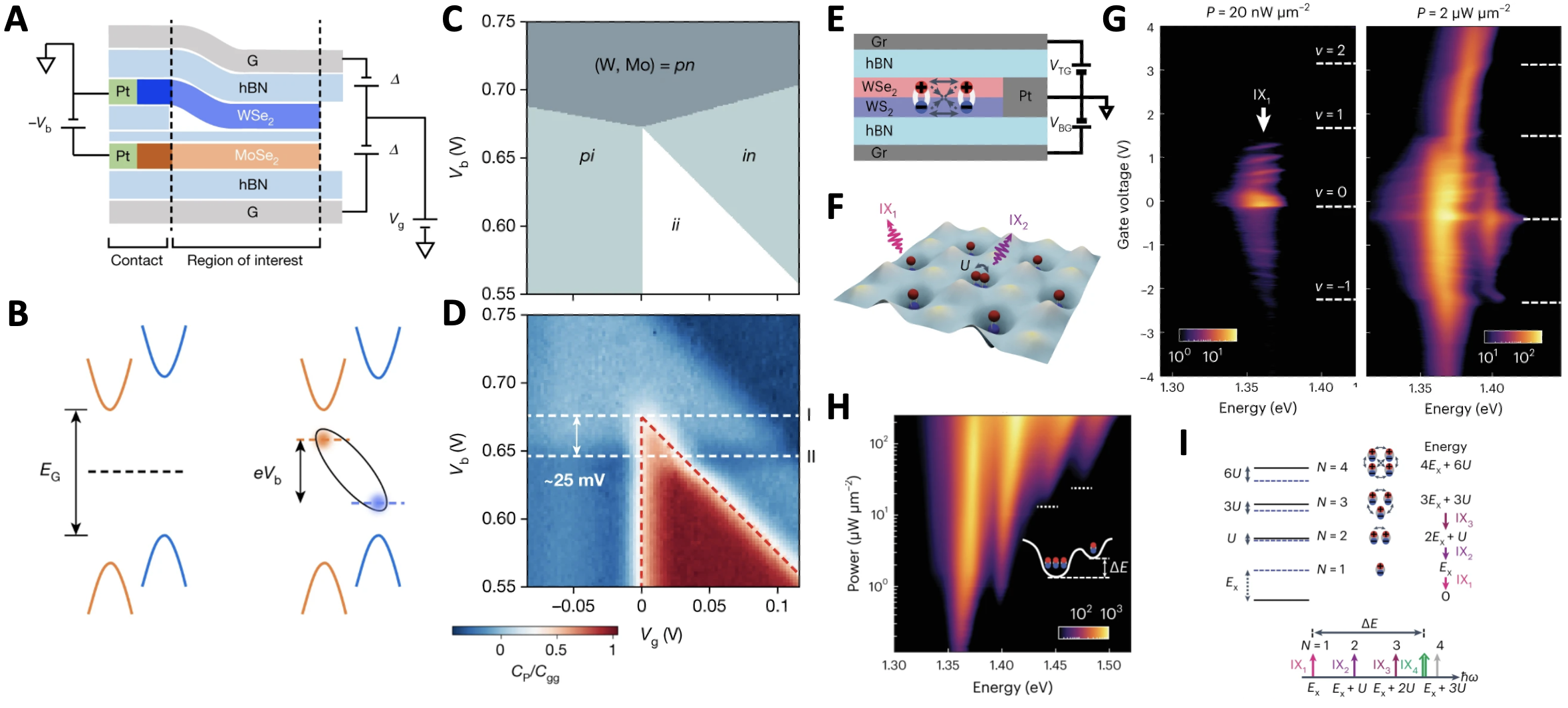}
   \end{center}
		\caption{Excitonic insulators in moir{\'e} superlattice structures [WSe$_2$/hBN/MoSe$_2$ (A-D), WSe$_2$/WS$_2$ (e-I)].\\
 (A) Schematic of dual-gated bilayer devices. Anti-symmetric gating $\Delta$ reduces the gap energy $E_G$. The symmetric gating $V_g$ tunes the electron and hole density difference. (B) Scheme of the type II band alignment of MoSe$_2$/WSe$_2$. A bias voltage $V_b$ reduces the charge gap. IXs can form. (C) Electrostatics simulation of the bilayer based on a parallel-plate capacitor model. Depending on bias and gate voltages, the bilayer can be in an intrinsic (i), positively (p) and negatively (n) doped regime. (D) Normalized penetration capacitance $C_P$/$C_ {gg}$ as a function of bias and gate voltages ($C_{gg}$ gate-to-gate capacitance). Charge incompressible region is marked by red dashed lines. The estimated exciton binding energy in the limit of zero exciton density is 25 meV. (measurements done at 15 K and $\Delta$ = 4.6 V). (E) Schematic of a dual-gated WSe$_2$/WS$_2$ hetero-bilayer. (F) Illustration of the moiré exciton lattice with one of the moiré unit cells with double occupancy.
(G) IX PL intensity as a function of gate voltage (doping) and emission energy for low (left) and high optical excitation intensity (right). Filling factors are labelled. With increasing excitation intensity, the PL peak $IX_2$ (34 meV above the ground state) appears. (H) IX PL intensity versus excitation power $P$. The threshold $P_{th}$ for the occurrence of $IX_3$ and $IX_4$ is marked by dashed lines. (I) Energy level diagram for a single moir{\'e} orbit occupied by multiple exciton dipoles. Solid (dashed) line indicates the scenario with (without) onsite dipole–dipole interaction. Expected PL spectrum from the dipole ladder with constant energy spacing between $IX_1$ and $IX_3$. $IX_4$ originates from the fourth exciton localized in the second moir{\'e} orbital and has hence lower energy. Adapted with permission from Ref. \cite{2021_Ma_exciton-ins_double} (A-D) and from Ref. \cite{park2023dipole} (E-I).}
	\label{Fig_5a}
\end{figure*}

First predicted theoretically in 2018 \cite{wu2018hubbard,naik2018ultraflatbands}, it has been experimentally shown that strong electronic correlations can arise in the flat-bands of TMD hetero-structures that arise due to the moir{\'e} superlattice structure \cite{tang2020simulation,regan2020mott,shimazaki2020strongly,xu2020correlated}.  In this scenario, the flat bands quench the kinetic energy of the charge carriers relative to their Coulomb interaction energy, and several stable charge ordered phases, designated as Wigner crystals and Mott insulators, are observed at multiple fractional fillings ($\nu$) of the moir{\'e} lattice. Selected examples of ordered states are shown in Figure \ref{Fig_5a}A.

Initially, the charge ordered states were optically probed using the dielectric response of IXs to the correlated electronic states \cite{tang2020simulation,regan2020mott,shimazaki2020strongly,xu2020correlated}. However, due to their large permanent dipole, IXs are not only highly tunable with applied electric field, but they are highly sensitive to their charge environment. Natural questions to ask then, is how do the correlated electronic phases affect the IX emission and can one use the trapped IX as a sensitive local probe of the electronic crystallization in the vicinity of the IX? In addition to use IXs as local probes, novel many-body ground states are formed via the interaction of moir{\'e} excitons and correlated electron lattices.

In a WSe$_2$/WS$_2$ moir{\'e} hetero-structure, Liu et al. investigated the PL from IXs as a function of fractional filling of the moir{\'e} lattice \cite{liu2021excitonic}. Abrupt changes in PL intensity and photon energy are observed at a number of different fractional fillings corresponding to correlated insulating states at $\nu$ = -2, -3/2, -8/7, -1, -1/3, -1/4, 1/4, 1/3, 2/5, 2/3, 6/7, 1, 5/4, 5/3, and 13/7. 
This modulation can arise because the insulating phases have reduced charge screening which renormalizes the band gap and exciton (or trion) binding energy. Further, the valley polarization is modulated by the charge ordered states. The degree of valley polarization is measured by spectrally integrating the PL intensity ratio (I$_{RR}$ - I$_{RL}$) / (I$_{RR}$ + I$_{RL}$), where the subscripts $RR$ and $RL$ indicate excitation-detection right- and left-handed circular polarization.  The degree of valley polarization clearly decreases at the transition from conducting to insulating states. The valley polarization is limited by intervalley scattering and thus dependent on the electron-hole exchange interaction, which increases when the screening effect is reduced. Hence, in the insulating states, the intervalley scattering is increased and the valley polarization reduced.

\subsection{Excitonic insulators in moir{\'e} structures}
IXs with at least one constituent residing in a moiré-flat band are prone to form correlated bosonic states similar to the fermionic Mott-insulator state. Experimental observations of excitonic insulator states in TMD hetero-structures are reported for various material combinations \cite{2023_Xiong_corr-ins-phasediagram, park2023dipole} including natural bilayers \cite{2022_Chen_insulator, 2022_Zhang_corr_IX_ins} and with ultrathin hBN separation layer \cite{2021_Ma_exciton-ins_double} as shown in Figure \ref{Fig_5a}(A). Strong onsite dipole-dipole interaction of excitons occupying the same moir{\'e} lattice is reported for WSe$_2$/WS$_2$ bilayers with the interaction between those IXs given by the Hubbard $U$ parameter \cite{park2023dipole} (see Figure \ref{Fig_5a} (E, F)). PL measurements reveal a dipole ladder with emission peak separation of around 34 meV as summarized in Figures \ref{Fig_5a} (G-I). The authors conclude  from such a large Hubbard parameter that in such systems exciton crystal phases can be possibly realized \cite{park2023dipole}. Signatures for the formation of an incompressible IX state in a MoS$_2$/hBN/WSe$_2$ hetero-structure has been deduced from capacitance measurements in Ref. \cite{2021_Ma_exciton-ins_double} (see Figures \ref{Fig_5a} (C-D)). An incompressible state in this context means that the chemical potential of the system increases discontinuously as a function of exciton density \cite{MacDonald_1994}. Similarly, signatures for the formation of an incompressible exciton state formed in WS$_2$/bilayer WSe$_2$ hetero-junction moir{\'e} superlattice for exciton filling factor $\nu$=1 have been observed by utilizing microwave impedance microscopy and differential reflectance spectroscopy $\Delta R / R$ \cite{2022_Chen_insulator}. Those signatures exhibit a peculiar temperature dependence and vanish above a critical temperature of about 90 K, suggesting the formation of an exciton insulator state of IXs at low temperatures \cite{2022_Chen_insulator}. Interestingly, Xiong et al. report the phase diagram for a mixed fermionic and bosonic correlated insulator for 60$^\circ$-aligned WSe$_2$/WS$_2$ \cite{2023_Xiong_corr-ins-phasediagram}. The excess electron density $n_{ex}$ and the exciton density $n_X$ were experimentally controlled and the mixed correlated insulator state has been observed along the line of $n_{tot} = n_{ex} + n{e} = 1$. Again, the exciton correlated insulator states are experimentally identified by their incompressible nature determined by a special type of optical pump-probe spectroscopy in analogy to the electrical capacitance measurements \cite{2023_Xiong_corr-ins-phasediagram}.

\subsection{Degenerate ensembles of mobile interlayer excitons}

At low temperatures and high exciton densities, thermalized ensembles of mobile IXs can be considered to be degenerate in vdW hetero-bilayers, as soon as the excitonic thermal de-Broglie wavelength exceeds the mutual distance between the IXs \cite{blatt_bose-einstein_1962,laikhtman_exciton_2009}. The corresponding phase diagram includes this  degeneracy phase with a predicted local superfluidity at a temperature below 10s of Kelvin  and a possible Berezinskii–Kosterlitz–Thouless transition to a phase with an expected macroscopic superfluidity at even lower temperatures \cite{fogler_high-temperature_2014}. At high exciton densities (above $\sim 10^{12}$ cm$^{-2}$), a Mott transition to a degenerate electron-hole Fermi gas is expected wherein fermionic interactions between the electrons and holes dominate \cite{fogler_high-temperature_2014, laikhtman_exciton_2009}. Recently, it was proposed that also for the degenerate phase both the fermionic and bosonic characteristics of the exciton ensembles need to be considered, particularly at the presence of phonons \cite{katzer_exciton-phonon_2023}.

\begin{figure*}[t]
	\begin{center}
		\includegraphics[scale=0.4]{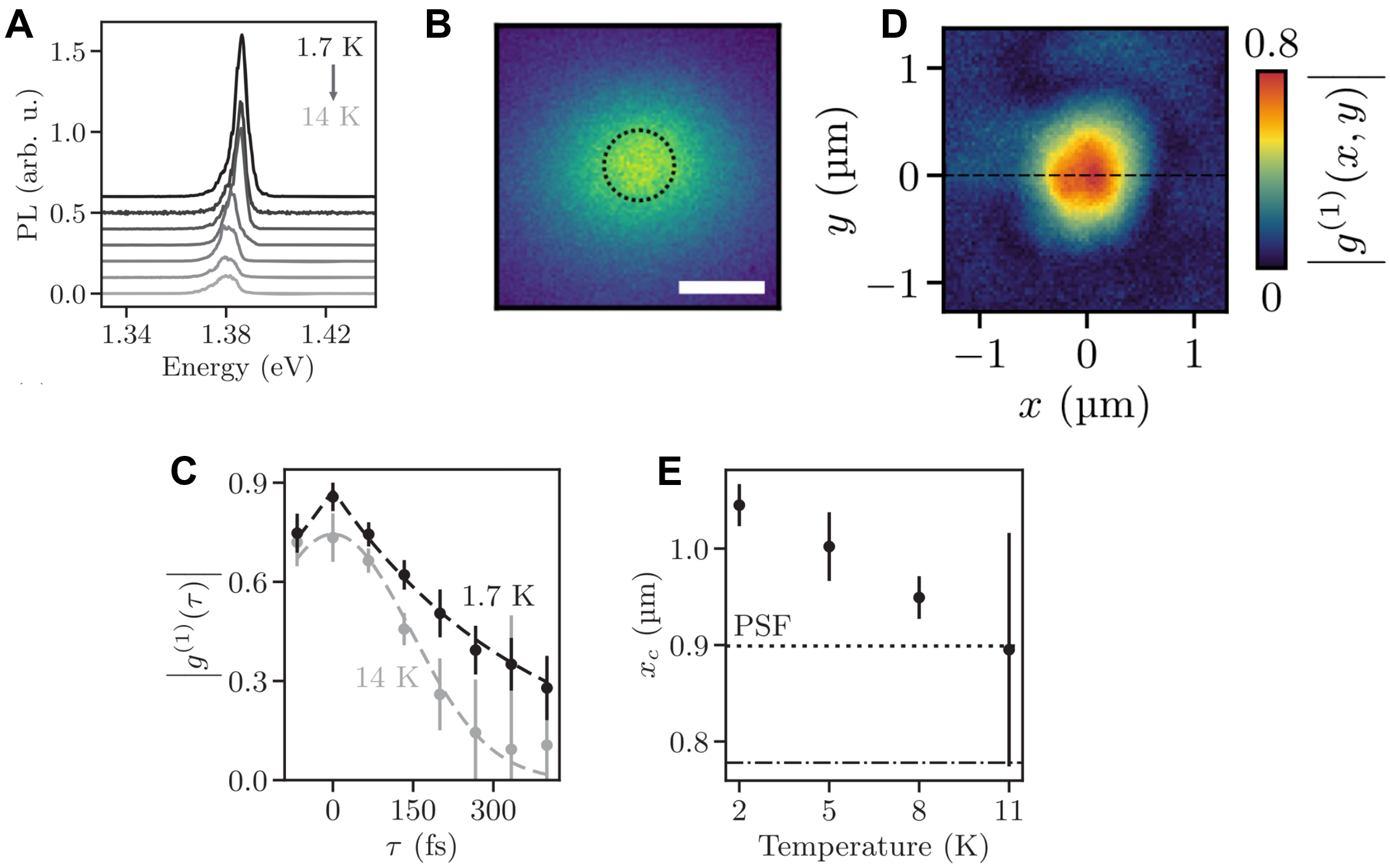}
	\end{center}
	\caption{Degenerate ensembles of mobile IXs and the coherence of their luminescence [MoSe$_2$/WSe$_2$ (A - E)].\\ (A) PL spectra from a MoSe$_2$/WSe$_2$ hetero-bilayer with h-BN encapsulation at the bath temperature ranging from 1.7 K to 14 K. The maximum at $\sim$1.38 $eV$ corresponds to IXs. (B) Corresponding spatial PL map  in a false-color plot at 1.7 K [green (blue) equals high (low) intensity], demonstrating an isotropic expansion of the IXs in the plane of the hetero-bilayer. Dotted circle highlights the point spread function (PSF) at the IX wavelength (1.38 $eV$ $\sim$ 0.9 $\mu$m). Scale bar, 1 $\mu$m. (C) Normalized first order correlation function $|g^{(1)}(\tau)|$ of the IX luminescence for 1.7 K and 14 K, with $\tau$ the time delay as given by a Michelson-Morley interferometer. (D) Spatial dependence of the normalized first order correlation function $|g^{(1)}(x,y)|$ at $\tau$ = 0. (E) Temperature dependence of the spatial coherence length $x_c$. Dotted and dashed-dotted lines represent the point spread function PSF at the emission wavelength [as in (B)] and the excitation energy (1.94 $eV$ $\sim$ 639 nm). Adapted with permission from Ref. \cite{troue_extended_2023}\\
	}
	\label{Fig_6a}
\end{figure*}

The coherence of the ground state of such many-body ensembles can be deduced from the temporal and spatial coherences as measured in luminescence experiments (see Figures \ref{Fig_6a}A and \cite{eisenstein_boseeinstein_2004, stern_exciton_2014, sigl_signatures_2020, troue_extended_2023, liu_quantum_2017, wang_evidence_2019, kogar_signatures_2017, alloing_evidence_2014, high_condensation_2012, griffin_bose-einstein_1995}). For such experiments, IXs are advantageous compared to intralayer excitons due to their relatively long lifetime of tens to hundreds of nanoseconds \cite{rivera_observation_2015,miller_long-lived_2017,hanbicki_double_2018,merkl_ultrafast_2019} and a permanent dipole moment \cite{kiemle2020control} (see above sections). The long lifetime allows performing time-resolved experiments, such that the IX ensembles can be considered to be thermalized before they emit a photon \cite{merkl_ultrafast_2019,troue_extended_2023}, while the out-of-plane dipole drives an in-plane expansion of the IXs. In turn, the IX mobility can be straight-forwardly “imaged” as soon as the spatial luminescence pattern exceeds the optical excitation area (Figures \ref{Fig_6a}B and section below, \cite{butov_magneto-optics_1999,voros_long-distance_2005,gartner_drift_2006,gartner_micropatterned_2007,vogele_density_2009,zimmermann_excitonexciton_2007, glazov_phonon_2019, wagner_nonclassical_2021,sun2021excitonic, datta_spatiotemporally_2022,peng_long-range_2022}). Recent experimental work on MoSe$_2$/WSe$_2$ hetero-bilayers suggests that IX ensembles particularly with a Lorentzian luminescence profile allow accessing coherent phases of dense and mobile IXs, consistent with the predicted quantum degeneracy at low temperature \cite{sigl_signatures_2020,troue_extended_2023}. In contrast, Gaussian luminescence profiles with possible sub-structures appear at higher temperatures and/or at the presence of additional localized excitons (cf. Figures \ref{Fig_6a}A and above sections).

In principle, luminescence experiments allow distinguishing whether bosonic or fermionic interactions dominate within the ensembles of mobile IXs: for an increasing exciton density, a decrease of the full-width at half maximum (FWHM) of homogeneously broadened luminescence spectra is understood as a signature for a dominating bosonic characteristic within the exciton ensembles \cite{katzer_exciton-phonon_2023}. A corresponding increase of the FWHM is less unambiguous, pointing towards fermionic interactions, but also towards the interaction with further particles, such as phonons \cite{selig_excitonic_2016, moody_intrinsic_2015, katzer_exciton-phonon_2023}. The temporal coherence of the luminescence can be accessed by the help of Michelson-Morley interferometers. The corresponding coherent part of the interferometers' signal is expressed as the normalized first order correlation function $|g^{(1)}(\tau)|$ \cite{loudon_quantum_2000}, with $\tau$ the time delay between the two optical paths within the interferometer (cf. Figures \ref{Fig_6a}C). Particularly, a bi-exponential decay of $|g^{(1)}(\tau)|$ as in Figures \ref{Fig_6a}C allows accessing the homogeneously broadened (Lorentzian) part of a luminescence spectrum (cf. Figures \ref{Fig_6a}A and \cite{loudon_quantum_2000}). Recent work suggests that degenerate ensembles of mobile IXs in MoSe$_2$/WSe$_2$ hetero-bilayers show such a coherent luminescence of “synchronized” emitters \cite{sigl_signatures_2020}, with the temporal coherence time, the FWHM of $|g^{(1)}(\tau)|$ being in the regime of 100s of fs. At present, this fast time scale is related to the emission process of the photons. At higher temperatures, when the IX ensembles are expected to be in the phase of a non-degenerate, classical gas, the corresponding $|g^{(1)}(\tau)|$ is reduced and it seems to follow a Gaussian profile (cf. Figures \ref{Fig_6a}C and \cite{troue_extended_2023}).

For measuring the spatial coherence of the IXs' luminescence, it is essential that the spatial pattern of the excitonic luminescence is isotropic in the plane of the hetero-bilayer (e.g. Figures \ref{Fig_6a}B) and that the point-spread function (PSF) of the utilized optical system both at the wavelength of the optical excitation and detection is thoroughly understood \cite{troue_extended_2023}. Then, with the help of a point-inverting Michelson-Morley interferometer \cite{savasta_quantum_2005, paik_interlayer_2019}, the spatial dependence of the  normalized first order correlation function $|g^{(1)}(x,y)|$ can be imaged, with $x$ and $y$ the coordinates within the reference frame of the hetero-bilayers (cf. Figures \ref{Fig_6a}D). Importantly, such spatial coherence experiments need to be performed again in a time-resolved manner, because only when the excitation laser is off for several 100s of fs to ps, laser-induced coherences as well as thermalization dynamics within the hetero-bilayers can be excluded to impact the IXs' luminescence \cite{troue_extended_2023}. The lateral FWHM of $|g^{(1)}(x,y)|$, e.g. as measured along the dashed line in Figure \ref{Fig_6a}D, gives access to the spatial coherence length $x_c$ of the IXs' luminescence. Recent experiments on MoSe$_2$/WSe$_2$ hetero-bilayers demonstrate that $x_c$ can exceed the PSFs of the optical imaging apparatus at the experimental temperature when the IXs can be theoretically considered to be degenerate (cf. Figure \ref{Fig_6a}E and \cite{troue_extended_2023}). Then, $x_c$ can be even equal to the overall expansion of the IXs ensembles \cite{troue_extended_2023}. Future studies, e.g. on laterally patterned hetero-bilayers \cite{schinner_confinement_2013, gartner_micropatterned_2007, kuznetsova_two-dimensional_2015, shanks_nanoscale_2021}, might reveal whether the predicted local or macroscopic superfluidity explains the extended spatial coherence of the luminescence. Moreover, for an unambiguous experimental evidence of a pure bosonic exciton condensation in the momentum space, back-focal plane imaging seems to be suitable, however, at temperatures significantly below 1K \cite{sigl_optical_2022}.

\subsection{Dynamics of interlayer excitons: formation and transport}

 As discussed, in most TMD-based hetero-bilayer type-II band alignment with efficient IX formation upon separation of electron and hole states in adjacent layers is commonly reported. Already in 2014, Hong et al. \cite{2014_Hong_ultrafast} experimentally demonstrated ultrafast interlayer charge transfer withing 50 fs of photo-excited carriers across the vdW gap in MoS$_2$/WS$_2$ hetero-bilayers. This process is surprisingly fast since the photoexcited carriers are supposed to live at the $K$-valley formed by transition-metal $d$-orbitals that are well localized within each layer such that tunneling over the vdW gap is required for charge transfer \cite{2022_Meneghini_ultrafast_charge-transfer-theory}. Combined experimental and theoretical effort established that this ultrafast charge transfer is mediated by efficient intervalley phonon scattering connecting the layer localized CB $K$-states with the strongly hybridized $\Sigma$-states facilitating ultrafast charge transfer \cite{2020_Madeo_formation-ARPES, 2021_Zimmermann_ultrafast_charge_transfer, 2022_Meneghini_ultrafast_charge-transfer-theory, 2022_Schmitt_formation_arpes, 2023_Bange_ultrafast}. By femtosecond momentum microscopy together with microscopic modelling, the authors in Ref. \cite{2023_Bange_ultrafast} showed that momentum indirect $\Sigma-K$ intralayer excitons and layer hybridized $h\Sigma-K$- IXs form via exciton-phonon scattering in WSe$_2$ monolayers and WSe$_2$/MoS$_2$ hetero-bilayers, respectively. Interestingly, the relative level alignment between the direct $K-K$- and the indirect $(h)\Sigma-K$-excitons matters in the radiative decay process. If the $\Sigma$-states are energetically lower compared to $K$ as in the hetero-bilayer, an exciton cascade transfers the $K-K$-exciton over the $h\Sigma-K$ into a true IX \cite{2023_Bange_ultrafast}. For a (nearly) degenerate alignment, as in WSe$_2$ monolayers, the exciton occupation decays predominantly radiatively via the bright $K-K$-exciton indicating the crucial role of level alignment that is prone to changes by internal and external stimuli such as embedding in vdW stacks that can e.g. impact interlyer hybridization or interlayer phonon coupling, dielectric environment, doping or strain and is hence is highly tunable \cite{2023_Bange_ultrafast, 2020_Niehues_strain-tuning}. The crucial role of the phonons in these ultrafast charge transfer processes has been directly confirmed from ultrafast electron diffraction visualizing the lattice dynamics in photoexcited WSe$_2$/WS$_2$ \cite{2023_Sood_ultrafast-bidirectional}.

Several partially competing impact factors underlie the transport properties of excitons species in vdW bilayer structures resulting in a wealth of different scenarios \cite{2023_Malic_transport-comment, 2024.Wietek, 2023_Tagarelli_hybrid_IX_transport, 2022_Knorr_moire-transport_theory}. Moir{\'e} superlattice potential modulation can result in nearly complete localized excitons while shallow twist and/or high IX densities foster tunneling between moir{\'e} sites resulting in hopping transport such that the diffusion length is highly dependent on materials combination and twist angle and can even reach a few microns in commensurate vdW bilayers \cite{2020_Choi_moiré-ix_diffusion, Yuan.2020, 2022_Knorr_moire-transport_theory, 2021_Li_IX-transport}. IXs with constituents localized in adjacent layers hold a permanent dipolar moment such that dipole-dipole interaction, i.e. repulsion modulates exciton transport in vdW bilayers and allows for long-range propagation \cite{2023_Tagarelli_hybrid_IX_transport, 2024.Wietek}. Large linear diffusion coefficients even at low IX densities have been shown for reconstructed, low-disorder MoSe$_2$/WSe$_2$ hetero-bilayers at cryogenic temperatures \cite{2024.Wietek}. Non-linear propagation in those structures arise from nearly equally contribution from exciton-exciton repulsion and annihilation \cite{2024.Wietek}. Layer hybridized IXs have a reduced out-of plane dipolar moment and dipolar interaction contributes less to exciton transport. Tagarelli et al. demonstrate that the degree of hybridization can be tuned by electric fields enabling control of IX transport in vdW hetero-bilayers \cite{2023_Tagarelli_hybrid_IX_transport}. Understanding and controlling of exciton transport in vdW bilayer structures is important for both the realization of efficient excitonic  and optoelectronic devices.

\section{Conclusion and Outlook}

We introduced the fundamental concepts of how the interplay of spins, valley degree of freedom, moir{\'e} superlattic formation, doping as well as exteral stimuli determine the properties of exciton species in homo- and hetero-bilayer TMD devices. The interplay of the different degrees of freedom is dictated by the precise details of the stacking of the bilayer TMDs such that their properties can be widely engineered. This review provides a snapshot of the state-of-the-art experimental demonstrations in the field, including the first devices that provide technological implementations that exploit the novel properties of IXs in moir{\'e} materials. Altogether, these results establish a strong foundation for a wide range of future fundamental studies and developing technologies. In the following, we provide a perspective on future challenges and a few worthwhile pursuits in this rapidly emerging topic.

One exciting prospect is exploiting the highly tunable properties of trapped IXs as quantum light sources. Because of the IXs' large permanent dipole, DC Stark tuning of the exciton energy over a wide range is facile. So far, the quantum nature of the emission has been demonstrated with individual IXs trapped via the intrinsic moir{\'e} potential, but an additional avenue to pursue is extrinsic trapping of IXs for quantum light sources, for example via localized strain or electrostatic potential via tunable gating \cite{wang2018electrical_cf, shanks2021nanoscale,thureja2022electrically}. The externally generated potentials for trapping can enable deterministic positioning and scalability for applications in quantum photonics. For these applications, nanophotonic structures will be required to achieve Purcell enhancement and increased light-matter interaction efficiencies.  An open question is how coherent the generated photons can be; can highly indistinguishable photons be generated in these platforms? Similarly, open questions exist about the potential to exploit the spin-valley degrees of freedom of trapped IXs to enable a coherent spin-photon interface. On the other hand, the intrinsically high density of precisely arranged quantum emitters in the moir{\'e} lattice provides a platform for investigations of sub- and super-radiant Dicke states with applications in quantum information science.    

From a many-body physics point of view, the long lifetimes and strong dipolar interactions of Bosonic neutral IXs in TMD moir{\'e} materials can yield correlated excitonic states, as described by the Bose-Hubbard model. Changing the band structure (determined by the material combinations), moir{\'e} period and exciton density provide access to a wide range of the Bose-Hubbard phase diagram in different regimes such as non-interacting Bose gas phase, superfluid phase, quadrupolar and dipolar exciton ensembles, Mott insulator phase or beyond the Mott transition an electron-hole plasma phase \cite{slobodkin2020quantum,lagoin2021key,erkensten2021exciton,wang2021diffusivity,gotting2022moire}. By adding a charge to the localized, neutral excitons via electrostatic gating, charged Fermi-polarons can be created, providing access to the Fermi-Hubbard model. Overall, a significant goal in the field is to realize designer many-body Hamiltonians in moir{\'e} superlattices as a new platform for quantum simulators. Employing IXs as probes of these systems remains a compelling concept. Moreover, for MoSe$_2$/WSe$_2$ hetero-bilayers, reported signatures of degenerate exciton ensembles with an extended lateral coherence \cite{sigl_signatures_2020, troue_extended_2023} suggest that hetero-bilayers also allow studying bosonic and fermionic correlations in mobile exciton ensembles. In addition to tuning the phase diagram by the stacking configuration during fabrication, an important target will be to \textit{in-situ} tune the interaction parameters. Two approaches can readily be pursued: applying hetero-strain to tune the geometry of the moir{\'e} lattice \cite{kogl2022moir} or controlling the relative twist angle between the bilayers \cite{2023_Inbar_quantum-twist-micro}. 

To date, most TMD moir{\'e} devices have consisted of only a handful of individual atomic layers; due to poor fabrication yield, a device consisting of a few TMD layers encapsulated in hBN layers and electronic gating with graphene layers is considered a ‘hero’ device in many laboratories. This limits the boundless opportunities to freely engineer and add to the many degrees of freedom for dipoles in multilayer devices. For instance, spin-layer locking adds an additional layer degree of freedom in the system while giant interlayer dipoles extending over more than two-layers can in principle be engineered. Further, multilayer devices with independently configurable hetero-interfaces yield exciting prospects such as multi-orbital moire superlattices or multiple moir{\'e}  periodicities of varying moir{\'e} potential strength. 

A final challenge, not solely confined to moir{\'e}  TMD devices but for the entire field of vdW moir{\'e} materials, is the development of improved fabrication techniques that take steps towards reproducible and homogeneous moir{\'e}  superlattices with deterministic periodicity. Current state-of-the-art fabrication techniques yield significant inhomogeneities within the moir{\'e} superlattice, including twist angle disorder and defects such as bubbles or wrinkles. Additionally,  substantial sample-to-sample variation presents reproducibility issues across different laboratories. Important questions to address include “How homogeneous can a moir{\'e} material be?” and “How does twist-angle disorder affect or determine the emergent physical properties of the moir{\'e}  material?” Further, increased reproducibility and moir{\'e} homogeneity are essential to realize more complex samples and devices with increasing number of layers and contacts. Finally, improved and reproducible low-resistance Ohmic contacts to TMDs would facilitate transport measurements on a regular basis, a crucial technique which can compliment optical investigations of moiré TMD devices.

\section{Authors Contributions}
All authors discussed the content of the manuscript and wrote it together. 

\section{Funding}
We gratefully acknowledge financial support from the German Science Foundation via Grants HO 3324/9-2, WU 637/4-2 and 7-1, the clusters of excellence MCQST (EXS-2111) and e-conversion (EXS2089), and the priority program 2244 (2DMP). M.B.-G. is supported by a Royal Society University Research Fellowship. B.D.G. is supported by a Chair in Emerging Technology from the Royal Academy of Engineering and the European Research Council (grant no. 725920)

\section{Competing Interests}
There is no conflict of interest.

\section{Data availability}
Included in cited original research papers.
%

\end{document}